\documentclass[draftcls,11pt,onecolumn,letterpaper]{IEEEtran}
\IEEEoverridecommandlockouts
\usepackage{enumerate}
\usepackage{float}
\usepackage{color}
\usepackage{graphicx}
\usepackage{amsmath} 
\usepackage{amssymb,algorithm}  
\usepackage{array}
\usepackage{makecell}
\usepackage [table]{xcolor}
\usepackage[center]{caption}
\usepackage{slashbox}
\usepackage{subcaption} 
\usepackage{setspace}
\usepackage{lineno}
\usepackage{caption}
\usepackage{verbatim}
\usepackage{algorithm}
\usepackage[noend]{algpseudocode}
\definecolor{lightgray}{gray}{0.9}

\newcommand{\beq}{\begin{equation}}
\newcommand{\eeq}{\end{equation}}

\newcommand{\bq}[1]{\begin{equation} \label{#1}}
\newcommand{\eq}{\end{equation}}
\newcommand{\bed}{\begin{displaymath}}
\newcommand{\eed}{\end{displaymath}}
\newcommand{\bea}{\bed\begin{array}{rl}}
\newcommand{\eea}{\end{array}\eed}

\newcommand{\barray}{\begin{array}{ll}}
\newcommand{\earray}{\end{array}}

\renewcommand{\tilde}{\widetilde}

\newcommand{\E}{\mathbf{E}}

\usepackage{changebar}

\newcommand\independent{\protect\mathpalette{\protect\independenT}{\perp}}
\def\independenT#1#2{\mathrel{\rlap{$#1#2$}\mkern2mu{#1#2}}}

\makeatletter
\def\BState{\State\hskip-\ALG@thistlm}
\makeatother

\title{RSSI-Based Distributed Self-Localization for Wireless Sensor Networks used in Precision Agriculture
}

\author{Pooyan~Abouzar, 
        David~G.~Michelson, 
        and~Maziyar~Hamdi\\
        Department of Electrical and Computer Engineering\\The
         University of British Columbia, Vancouver, BC, Canada \\E-mail: \{pooyanab, davem, maziyarh\}@ece.ubc.ca}
\begin{document}
\maketitle
\vspace{-16mm}
\begin{abstract}
\setstretch{1.35}
\noindent
Node localization algorithms that can be easily integrated into deployed wireless sensor networks (WSNs) and which run seamlessly with proprietary lower layer communication protocols running on off-the-shelf modules can help operators of large farms and orchards avoid the difficulty, cost and/or time involved with manual or satellite-based node localization techniques. Even though the state-of-the-art node localization algorithms can achieve low error rates using distributed techniques such as belief propagation (BP), they are not well suited to WSNs deployed for precision agriculture applications with large number of nodes, few number of landmarks and lack real time update capability. The algorithm proposed here is designed for applications such as pest control and irrigation in large farms and orchards where greater power efficiency and scalability are required but location accuracy requirements are less demanding. Our algorithm uses received signal strength indicator (RSSI) values to estimate the distribution of distance between nodes then updates the location probability mass function (pmf) of nodes in a distributed manner. At every time step, the most recently communicated path loss samples and location prior pmf received from neighbouring nodes is sufficient for nodes with unknown location to update their location pmf. This renders the algorithm recursive, hence results in lower computational complexity at each time step. We propose a particular realization of the method in which only one node multicasts at each time step and neighbouring nodes update their location pmf conditioned on all communicated samples over previous time steps. This is highly compatible with realistic WSN deployments, e.g., ZigBee which are based upon the ad hoc on-demand distance vector (AODV) where nodes flood route request (RREQ) and route reply (RREP) packets. Further, beacon signals transmitted during the network formation and routing table formulation stage can provide the RSSI information required by the localization algorithm.
\end{abstract}
\section{INTRODUCTION}\label{introduction}
\vspace{-2mm}
\noindent With the advent of short range wireless technologies and standards in late 1990's variety of wireless localization techniques for indoor and outdoor applications have been developed. Wide range of indoor localization techniques have emerged based on camera, infrared, wireless local area network (WLAN), ultra wide band (UWB), Bluetooth, and radio-frequency identification (RFID) \cite{mautz} whereas global positioning system (GPS) technology revolutionized outdoor localization. Even though GPS-based localization techniques are attractive in terms of accuracy, their impaired coverage in metropolitan environments and lack of cost-effective scalable solutions sparked emergence of IEEE802.15.4/ZigBee RSSI-based localization algorithms. These techniques have advantage over Bluetooth, UWB and Wi-Fi due to their energy efficiency and capability to support high-range communication and mesh networking \cite{802154}. 

Localization techniques have been developed for different types of applications and are compared in terms of accuracy, coverage, cost, responsiveness and adaptiveness to environmental changes \cite{locmetrics,allen}. While some techniques such as laser and camera-based technologies are highly accurate and scalable in terms of coverage, they are usually too expensive to use for large environment applications. Particularly for large scale outdoor applications such as agricultural environments, a cost-effective, scalable and fast localization technique which is robust against seasonal environmental variations, e.g., growing season changes, is needed. On the other hand, accuracy requirements are usually looser because of relatively high inter-node distances which correspond to distance correlation of the measured features.   

One of the rapidly growing WSN areas for outdoor environments is precision agriculture which enhances crop management and yield through sophisticated management of soil, water resources and applied inputs \cite{Cassman}. WSNs are deployed to improve spatial data collection, precision irrigation, variable-rate technology and supplying data to farmers \cite{PAsurvey}. This requires sampling of critical features such as soil pH, moisture, electrical conductivity in addition to deployment of actuators to trigger wide variety of processes varying from drip irrigation to pest management, e.g., mating disruption. In order to provide meaningful feature maps that improve resource management and decision making, it is critical to be aware of location of the sensors that have generated data. Loose accuracy requirements, beside the cost involved with equipping all sensors with GPS, raise the need for localization algorithms which are low cost, and are compatible with commercial off-the-shelf (COTS) transceiver modules.
  
Anchor-based localization algorithms make use of landmarks or anchor nodes to help localizing unknown nodes \cite{localizationsurvey} and are divided into range-based and range-free techniques. Range-free algorithms on the other hand, only take advantage of the connectivity information \cite{Shang}, i.e., whether nodes are within the communication range of each other whereas range-based algorithms exploit time of arrival (TOA), angle of arrival (AOA) or RSSI to estimate the distance between nodes, so called inter-node distances. RSSI-based techniques are attractive in the sense that no additional hardware is required in order to make the distance estimation \cite{Wang}. Further even though AOA and TOA-based techniques are more precise, they are more complex in the sense that the former requires multiple antennas to detect signal arriving from different directions whereas the latter demands a large bandwidth for better multi path resolution.

This work is a probabilistic distributed and range-based localization technique for static WSNs based on RSSI samples, Bayesian model for information aggregation and particularly suited to precision agriculture applications. Most of the probabilistic distributed localization techniques work based on marginalization over a Markov random field (MRF) where joint distribution of nodes location $\left\{x_{1},x_{2},\ldots x_{n}\right\}$ based on noisy distance measurements between pairs of nodes $\left\{d_{ij}\right\}$ is expressed as multiplication of node and pairwise potentials, $P(x_{1},\ldots,x_{n}|\{ d_{ij}\}) \propto \underset{(i,j)}{\prod}{P(d_{ij}|x_{i},x_{j})} \underset{j}{\prod}{P(x_{j})}$ \cite{nonparametricbp}. Message passing algorithms such as belief propagation (BP), nonparametric belief propagation (NBP) and their variants are proposed to estimate the marginalization, hence location of each node \cite{nonparametricbp,SavicVariantsBP,UnderstandingBP,DistributedMP}. 
BP-based techniques are vulnerable to loopy graphs which cause them either not to converge at all or converge only under specific circumstances in terms of number of loops \cite{loopybelief}. Therefore these techniques have been mostly used for the scenarios where a few slowly moving or static nodes along with relatively high number of anchors, and all equipped with short range transmitters, render the statistical graph spanning tree or have few number of loops. Another shortcoming of these techniques is the need for global information from distance measurements to be available so that statistical graph is formed and algorithm could start to run. These two reasons lead to the fact that even though a relatively high accuracy is achieved with these techniques, remarkable amount of communication overhead, at least $O(n)$ depending on the technique, is required to form the spanning tree or statistical graph using multi-hop communications. The second issue is addressed in \cite{peng}, where nodes only exchange information with their single hop neighbours, however the communication and computation overhead required for making spanning trees with landmarks designated as root and other nodes keeping track of paths still holds since the procedure demands for independence of paths that arrive at the updating node. In contrast, in precision agriculture applications, relatively high number of connected unknown nodes resulted from high transmit power level, and underlying IEEE802.15.4 WSNs which work in conjunction with route discovery phase of AODV, call for a real-time algorithm which relies on local single hop information and is not susceptible to loops in the network.          

Our work is similar to \cite{peng} in the sense that nodes only communicate with their single hop neighbours and update their location in a real time manner rather than having to make the statistical graph using multi hop communications as in MRF-based approaches. However our algorithm needs no initialization in terms of spanning tree construction or having to start from a specific node or landmark in the field. In other words, the proposed technique is well positioned to address self-localization in do it yourself (DIY) networks which run ZigBee or other proprietary mesh networking protocols on top of IEEE802.15.4 specifications. The reason behind this is that the algorithm starts to work in conjunction with route discovery phase of AODV-based routing protocols such as ZigBee where route request packet (RREQ) originated from an arbitrary source node is flooded in the entire network. We derive a closed-form recursive relationship for Bayesian update of nodes location at a time step during which one or multiple path loss samples are generated therefore call it a Bayesian model for information aggregation. We prove that the location constraint resulted from a generated path loss sample is in fact convolution of path loss likelihood and the most recent location estimation of the generating node. Realistic independence assumptions, resulted from our measurements, are made to prove that location constraints resulted from dependent paths (loop forming paths)  multiply. This makes the algorithm faster by eliminating spanning tree construction, intermediate node tracking, and also making use of constraints resulted from the paths traversed by flooding RREQ packets, whereas algorithm's robustness against loops is verified by extensive simulations.

Since our goal is to devise an algorithm that can work in conjunction with COTS transceiver modules, we characterize path loss at 2.45~GHz industrial, scientific and medical (ISM) band. Based on our measurements in apple orchards, log-normal path loss model is proposed for high density apple orchards and for different transmitter (Tx) and receiver (Rx) antenna heights. Further, Rx was placed below tree height whereas Tx was fixed below and above the tree height. In the rest of this paper, these two antenna height modes are called below and above canopy level respectively. The path loss data was collected during three measurement campaigns throughout two consecutive summer seasons. 

The remainder of this paper is organized as follows: In Section \ref{ProblemFormulation}, we formulate the localization problem, define the notations, include a brief summary from our measurement campaigns and explain the path loss model along with path loss likelihood function conditioned on node locations. In Section \ref{IterativeAlgorithm}, we devise a recursive solution to the problem stated in Section \ref{ProblemFormulation} and propose a specific implementation of this solution based on nodes multicasting in TDMA manner. Finally we proceed with simulations and evaluation of our algorithm in Section \ref{results} and wrap up the paper with conclusion in Section \ref{conclusion}.  

\section{The Localization Problem and Path Loss Likelihood Function}
\label{ProblemFormulation}
\noindent As stated in Introduction, pinpoint localization accuracy is not required for  precision agriculture applications such as pest control since knowing  approximate location of originating sensors suffices to trigger the relevant actuators. Accordingly, we define the localization problem in a discrete manner which means that the agricultural field is divided into smaller square cells and  location of each unknown node is determined as centroid of one of the cells the field is divided into. The precision of the algorithm is adjustable via number of grid cells inside the field, however precision flattens once grid resolution exceeds a threshold. Formulation of the localization problem based on aggregated path loss samples from neighbour nodes is discussed in Section \ref{problemformulation} and path loss model for orchard environments is explained briefly in Section \ref{channelmodel}.       
  
\subsection{Problem Formulation}  
\label{problemformulation}
\noindent Let $S=\{S_{1},\ldots,S_{N}\}$ be a set of sensors randomly scattered in a square field which is divided into $m\times m$ square cells with equal areas, and $\Omega=\{1,2,\ldots,m^{2}\}$ be the sample space of all possible cell coordinates. Our objective is to make use of inter-node communications and find the grid cell each node is located in. In the following, we introduce the notations and formalize the localization problem. 

Without loss of generality, let the first $n_{a}$ nodes be landmarks $S_{l}=\{S_{1},\ldots,S_{n_{a}}\}$, and unknown nodes be represented by $S_{u}=\{S_{n_{a}+1},\ldots,S_{N}\}$ while $y^{l}_{ij}$ is a path loss sample or average of multiple path loss samples that  $S_{j}$ collects from $S_{i}$ at $l$th time step. Note that in general, multiple samples could be collected in case each calculation time step is made up of multiple communication time slots. 
Let $\Theta_{k}$ denote vector of path loss samples which have been communicated between pairs of connected nodes during the first $k$ time steps and let $Y^{(k)}_{j}$ represent vector of all path loss samples that $S_{j}$ has collected from its neighbour nodes with index set $N_{j}$ at $k$-th time step, 
\begin{equation}
\left\{\begin{split}&\Theta_{k}=(y^{l}_{ij})_{\tiny{\begin{matrix} l=0:k \\ 1 \leq j\leq N, i \in N_{j} \end{matrix}}} \\
&Y^{(k)}_{j}=(y^{k}_{ij})_{i \in N_{j}} \end{split}  \cdot \right.
\end{equation}

Note that $y^{k}_{mj}$ is not available in case $S_{j}$ has not collected any sample from $S_{m}$ at $k$-th time step. Let $\bold{\bold{\tilde{X}^{(k)}_{j}}}$ be a random variable defined over $\Omega$ representing location estimation of $S_{j}$ at $k$-th time step. Considering that we are looking to estimate location of $S_{j}$ at $M$th time step based on previous aggregated data $\Theta_{M}$, 
\begin{equation}
\label{problem}
\tilde{x_{j}}=\underset{x_{j}}{\operatorname{argmax}}{[P(\bold{\tilde{X}^{(M)}_{j}}=x_{j}\big| \Theta_{M})]},
\end{equation}
where $P(\cdot)$ is the probability function and $\underset{x}{\operatorname{argmax}}{[f(x)]}$ is the set of points $x$ for which $f(x)$ attains its largest value. In the remainder of this section, path loss model for agricultural environment which is the key to generate $y^{l}_{ij}$ samples, $\Theta_{k}$ and $Y^{(k)}_{j}$ is explained. Consequently we derive the path loss likelihood function that underpins the recursive algorithm described in Section \ref{IterativeAlgorithm}. Moreover, we derive likelihood of  $y^{l}_{ij}$ given that $S_{i}$ and $S_{j}$ are estimated to be located at $x_{i}$ and $x_{j}$ respectively, i.e., $P(y^{k}_{ij} \big| \bold{\tilde{X}^{(k)}_{j}}=x_{j},\bold{\tilde{X}^{(k)}_{i}}=x_{i})$.   
\subsection{A Representative Path Loss Model for Orchard Environments}
\label{channelmodel}
\noindent In this section, we explain the path loss model resulted from our measurement campaigns in apple orchards located at Keremeos, BC, Canada. This underlies the work in Section \ref{PLLikelihood} which explains derivation of path loss likelihood function expressing path loss distribution conditioned on Tx and Rx locations. 

\noindent There is an extensive literature on path loss models for forests and agricultural environments. It is claimed that log-distance path loss model provides a good fit to the measured path loss in vegetated environments \cite{lognormal1,lognormal2, lognormal3},
\begin{equation}
PL[dB]=PL_{0}+10n\log(\frac{d}{d_{0}})+X_{\sigma},
\end{equation}
where $X_{\sigma}$ is a zero-mean normal random variable with standard deviation $\sigma$, $X_{\sigma}\sim N(0,\sigma)$, whereas $PL_{0}$ represents path loss at reference distance $d_{0}$ and $n$ denotes path loss exponent for the specific case of study. 

We carried out the measurements in Dawson orchards at Keremeos, Okanagan, British Columbia. Measurements were conducted in a 6~hectare (ha) orchard consisting of apple tree rows divided into standard and high density in terms of vegetation and canopy density with trees being approximately 3~m high. We use the path loss data collected from four directions of along, cross, $30^{\circ}$, $45^{\circ}$ and $60^{\circ}$ with respect to tree rows, using different transmitter (Tx) and receiver (Rx) antenna heights. Further, we conducted measurements with Tx at 2.5~m (below canopy level) and 4~m (above canopy level) heights and Rx at 2.5~m. This setup is compatible with realistic WSN deployment scenarios where gateways, responsible for aggregating data of their neighbouring sensors, are mounted above canopy whereas sensors and actuators are placed inside the canopy. As localization is concerned, gateways which have better line of sight (LOS) are equipped with GPS to play the landmark role. The measurements were conducted throughout three different measurement campaigns, seven days combined and spread across two summer seasons.

\begin{figure}[h!]
\begin{subfigure}[b]{0.5\textwidth}
\includegraphics[width=\textwidth,height=7cm]{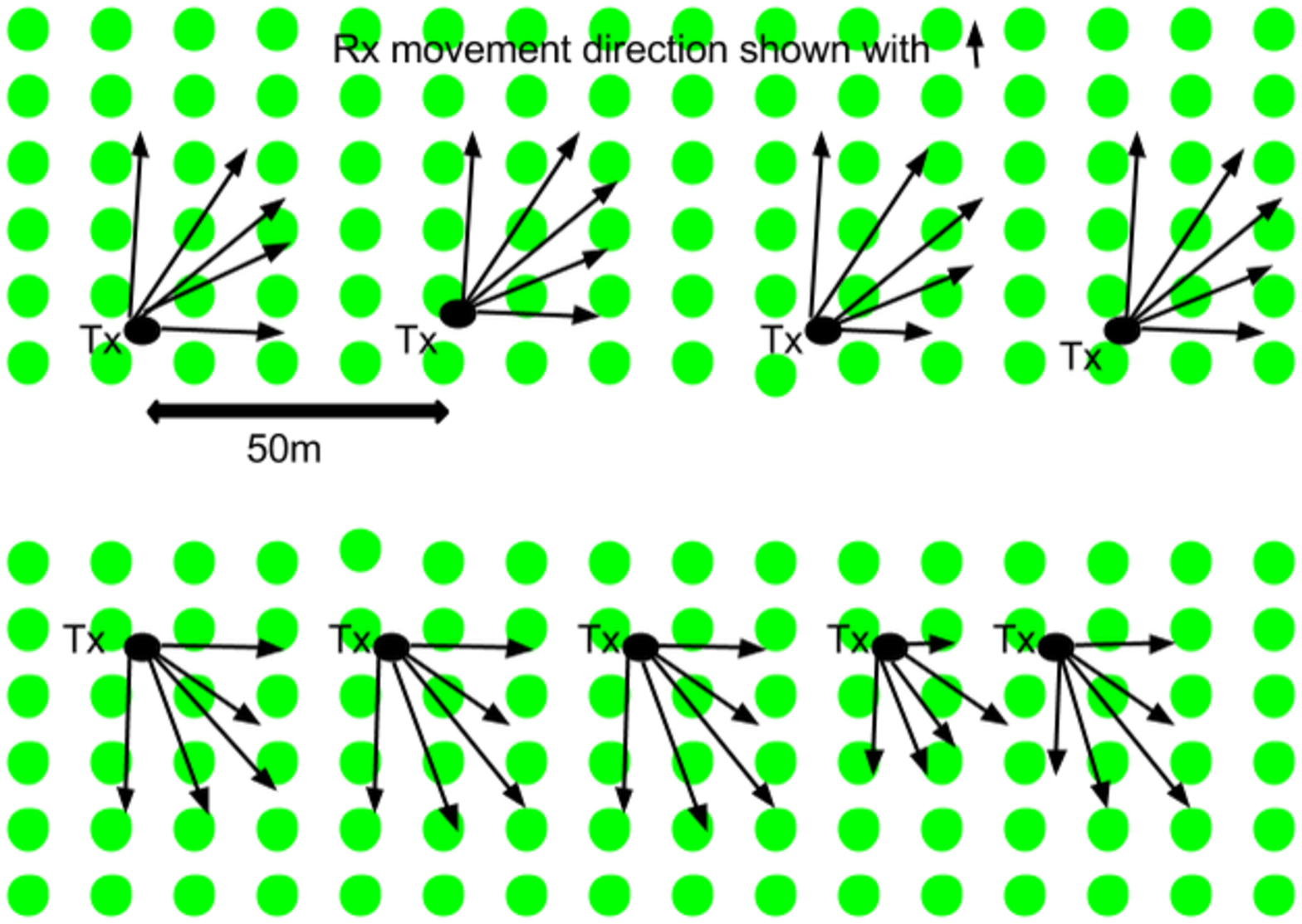}
\caption{}
\end{subfigure}
\begin{subfigure}[b]{0.5\textwidth}
\includegraphics[width=\textwidth,height=7cm]{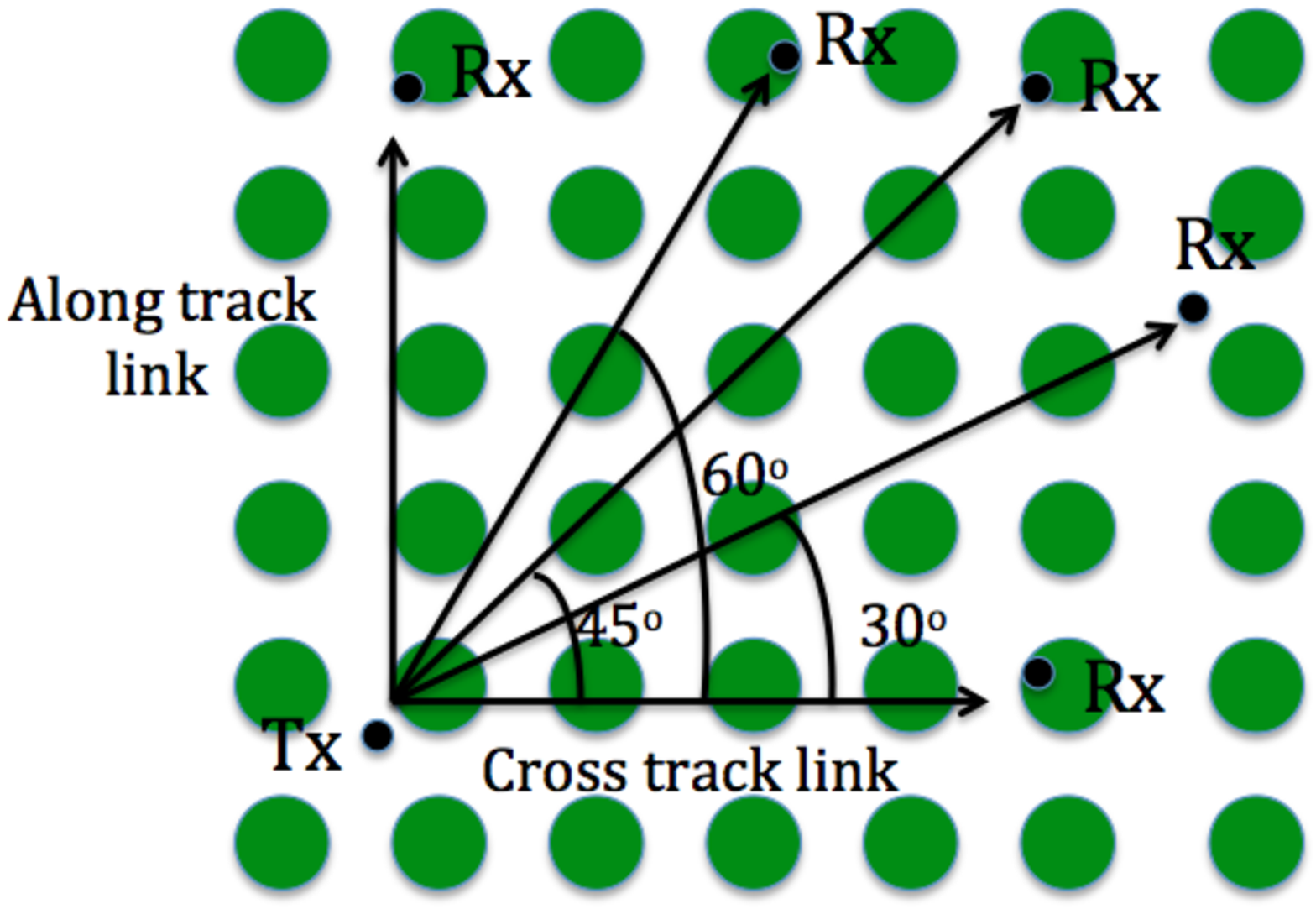}
\caption{}
\end{subfigure}
\caption{9 measurement scenarios inside the orchard is illustrated; Transmitter antenna was moved 50~m across the rows to form a new scenario whereas Rx was moved along four different directions of along, cross, $30^{\circ}$, $45^{\circ}$ and $60^{\circ}$ for each scenario and path loss samples were collected through 0-100~m range and at $\approx10~m$ apart points. Rx antenna was placed at 2.5~m elevation (0.5~m below tree height) while Tx antenna height was at 2.5~m and 4~m elevation (1~m above canopy level). }
 \label{measscenarios} 
\end{figure}

Measurements were done in approximate range of 0-100~m at points which are approximately 10~m apart from each other at 9 different parts of the orchard along four directions illustrated in Figure \ref{measscenarios}. Our equipment on the transmitter side, are an Agilent E8267D vector signal generator (VSG) feeding a 2.45~GHz omnidirectional dipole antenna with 5 multi-tones (5~MHz apart from each other) through a ZVA-213 power amplifier which provides +23~dBm as the antenna input. Whereas on the receiver side, a Toshiba laptop which runs MATLAB and Agilent connection expert, specialized proprietary software for connecting computer to Agilent spectrum analyzer, is connected to a N9342C handheld spectrum analyzer (HSA) via a LAN cable. Extra losses and gains resulted from cables, connectors and antennas at both Tx and Rx sides have been taken into account for calibration. 
 
\begin{figure}[!htb]
\centering 
\includegraphics[width=3.5in,height=6cm]{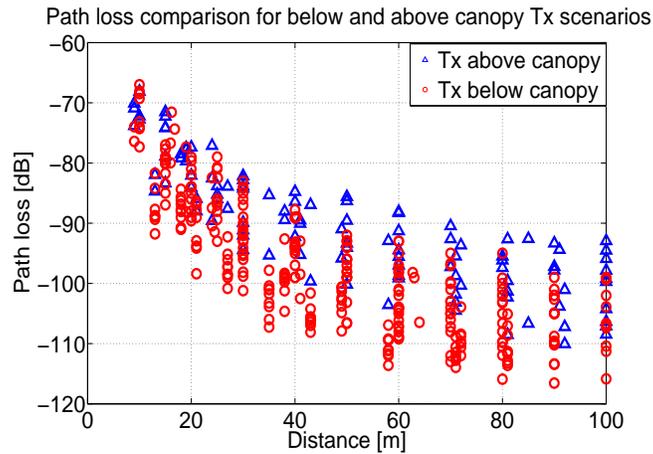}
\caption{Path loss samples for below and above canopy Tx level at 2.45~GHz collected from three measurement campaigns; The difference between the two above and below canopy modes, which is due to more line of sight (LOS) between Tx and Rx in the above canopy case, could be seen.}
\label{abovebelowcomparison}
\end{figure}

The summary of path loss statistics along with statistical measure $R^{2}$, which indicates how well data fits the log-distance model, and 95\% confidence interval (CI) for $PL_{0}$ and $n$ are expressed in Table \ref{pathlosstable}, whereas path loss samples for two modes are illustrated in Figure \ref{abovebelowcomparison}. Note that gateway-to-node and node-to-node communications comply with above and below canopy level Tx modes respectively.  
\begin{table}[!htb]%
\caption{Path Loss Model characteristics for above and below canopy level modes}
\label{pathlosstable}\centering %
\rowcolors{1}{}{lightgray}
\begin {tabular}{ccccccc}
Mode & $n$  & $PL_{0}[dB]$ & $\sigma [dB]$ & $R^{2}$& 95\% CI for n & 95\% CI for $PL_{0}$ \\
\hline
2.45~GHz-Tx below canopy level& 3.61& 75 & 5.27 & 0.74&3.36-3.86&71-79  \\  
2.45~GHz-Tx above canopy level& 2.91 & 72& 4.14& 0.78&2.60-3.22&67-77 \\ 
\end{tabular} 
\end{table}

\begin{figure}[!htb]
\centering 
\includegraphics[width=3.5in]{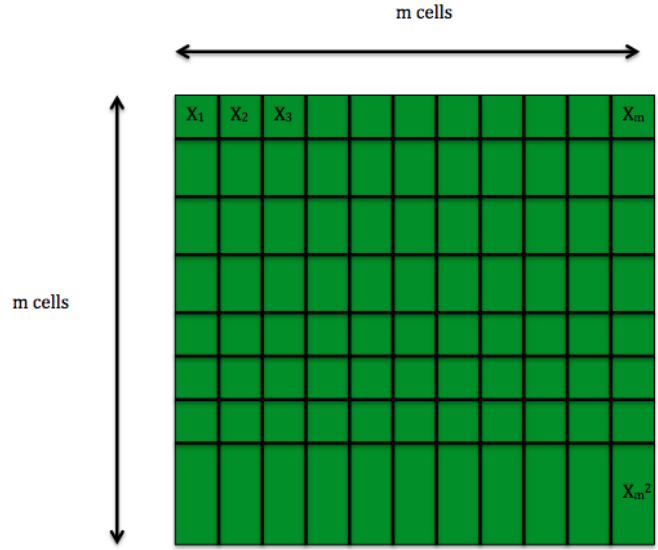}
\label{algorithmexample}
\caption{Location pmf of unknown nodes is updated recursively. Agricultural field is divided into $m \times m$ cells with equal area and probability of an unknown node being located inside each cell is calculated based on recently aggregated path loss samples and prior location pmf of connected nodes.}
\end{figure}
\subsection{Path Loss Likelihood Function}
\label{PLLikelihood}
\noindent In this part, we derive likelihood function $P(y^{k}_{ij} \big| \bold{\tilde{X}^{(k)}_{j}}=x_{j},\bold{\tilde{X}^{(k)}_{i}}=x_{i})$ which is a key component of the algorithm we propose in the next section since it relates path loss values to inter-node distances. 
 Assuming log-distance path loss model as discussed in Section \ref{channelmodel} and taking a random point on the field into account, the probability of path loss sample $y^{k}_{ij}$ falling in the range $[pl_{ij}-\frac{\Delta}{2},pl_{ij}+\frac{\Delta}{2}]$ with $\Delta<<pl_{ij}$ and when $S_{j}$ is located at distance $d_{ij}$ from $S_{i}$ is calculated by

\begin{equation}
\label{logdisprobability}
P\bigg(pl_{ij}-\frac{\Delta}{2}<y^{k}_{ij}<pl_{ij}+\frac{\Delta}{2}\bigg{|}D=d_{ij}\bigg)= \frac{C\Delta}{\sqrt{2\pi \sigma^{2}}}e^{-\frac{(pl_{ij}-\overline{PL(d_{ij})})}{2\sigma^{2}}},
\end{equation}
where $\overline{PL(d)}=PL_{0}+10n\log(\frac{d}{d_{0}})$ and $C$ is the normalization constant. Based on (\ref{logdisprobability}), and the fact that each pair $(x_{i},x_{j})$ translates into the corresponding distance $d_{ij}$, sensor $S_{j}$ calculates $P(y^{k}_{ij}=pl_{ij}|\mathbf{\tilde{X}^{(i)}_{j}}=x_{j},\mathbf{\tilde{X}^{(i-1)}_{i}}=x_{i})$, $\forall x_{i},x_{j} \in \Omega$.  Further in practice, in order to approximate the above conditional probability, we collect amplitude of the normal distribution with mean $\overline{PL(d_{ij})}$ and standard deviation $\sigma$ in the range $\left[\overline{PL(d_{ij})}-3\sigma , \overline{PL(d_{ij})}+3\sigma \right]$ at 1~dB steps and normalize the values so that they sum up to one. Note that the proposed path loss model in Section \ref{channelmodel} is used to derive the path loss likelihood function and also to generate random path loss samples in our simulations in Section \ref{results}.   

\section{Localization Algorithm For Precision Agriculture Applications}
\label{IterativeAlgorithm}
\noindent In this section, we derive an algorithm for the problem stated in (\ref{problem}) which works based on Bayesian model for information aggregation. Therefore, our objective is to derive a recursive expression for $P(\bold{\bold{\tilde{X}^{(k)}_{j}}}=x_{j}\big| \Theta_{k})$ that explains how location pmf is updated once information is aggregating in the network or in other words, the most recent evidence, RSSI sample, is collected. In Section \ref{generalalgo}, we first solve the problem for general case where at each calculation time step, arbitrary amount of information or number of packets, between one or multiple pairs of nodes is exchanged. In Section \ref{generalalgo}, we proceed with the special case which is more compatible with route discovery phase of AODV-based routing protocols such as ZigBee. This is the algorithm we have simulated in Section \ref{results}.   

\subsection{General Case}
\label{generalalgo}
\noindent According to the notation explanation in Section \ref{ProblemFormulation} and assuming that at each time step, $S_{j}$ updates its location pmf only based on the samples it has received from single hop neighbours, i.e., not  samples communicated between other pairs of nodes, 
\begin{equation}
\label{iteration1}
P(\bold{\bold{\tilde{X}^{(k)}_{j}}}=x_{j}\big| \Theta_{k})=P(\bold{\bold{\tilde{X}^{(k)}_{j}}}=x_{j}\big| \Theta_{k-1}, Y^{(k)}_{j}).
\end{equation}
Based on the fact that  $\Theta_{k-1} \independent Y^{(k)}_{j}$,
\begin{equation}
\label{iteration2}
P(\bold{\bold{\tilde{X}^{(k)}_{j}}}=x_{j}\big| \Theta_{k-1},Y^{(k)}_{j}) \propto P(Y^{(k)}_{j} \big | {\bold{\tilde{X}^{(k)}_{j}}}=x_{j}, \Theta_{k-1})P({\bold{\tilde{X}^{(k)}_{j}}}=x_{j} \big| \Theta_{k-1}).
\end{equation}
Let us recall that in general each calculation time step could be made up of several communication time slots therefore we have used $Y^{(k)}_{j}$ which are the path loss samples $S_{j}$ collects from one neighbour or a set of neighbours at $k$-th time step. Rephrasing (\ref{iteration1}) yields the recursive form, 
\begin{equation}
\label{iteration3}
P(\bold{\bold{\tilde{X}^{(k)}_{j}}}=x_{j}\big| \Theta_{k})\propto P(Y^{(k)}_{j} \big | {\bold{\tilde{X}^{(k)}_{j}}}=x_{j}, \Theta_{k-1})P({\bold{\tilde{X}^{(k)}_{j}}}=x_{j} \big| \Theta_{k-1}).
\end{equation}
We then simplify  $P(Y^{(k)}_{j} \big | {\bold{\tilde{X}^{(k)}_{j}}}=x_{j}, \Theta_{k-1})$ in the right-hand side of (\ref{iteration3}). Letting $\independent$ denote statistical independence and assuming that
\begin{equation}
\label{independentsamples}
 y^{k}_{ij} \independent \big(y^{k}_{mj} \big| \bold{\tilde{X}^{(k)}_{j}}, \Theta_{k-1}\big) \forall i,m \in N_{j} \, , 
 \end{equation}
First term on the right-hand side of (\ref{iteration3}) could be written as 
\begin{equation}
\label{iteration4}
P(Y^{(k)}_{j} \big | {\bold{\tilde{X}^{(k)}_{j}}}=x_{j}, \Theta_{k-1})=\prod_{i \in N_{j}}{P(y^{k}_{ij}\big | \bold{\tilde{X}^{(k)}_{j}}=x_{j}, \Theta_{k-1})}. 
\end{equation}
Our measurements followed by the procedure in \cite{correlation1} verify the assumption in (\ref{independentsamples}). Further our measurements show that shadowing correlation between links in the vegetated environment, which is the case of our study, is very low (below 0.1). This is reasonable due to long links we are dealing with which are $\approx 50~m$ for pest management applications. Due to lack of space and irrelevance to the main topic, we spare reader details on shadowing correlation calculation. 

Based on conditional expectation rule, we simplify the right-hand side of (\ref{iteration4}), 
\begin{align}
\begin{split}
\label{iteration5}
&P(y^{k}_{ij}\big | \bold{\tilde{X}^{(k)}_{j}}=x_{j}, \Theta_{k-1})=\sum_{x_{i}}{P(y^{k}_{ij} \big| \bold{\tilde{X}^{(k)}_{j}}=x_{j},\bold{\tilde{X}^{(k)}_{i}}=x_{i}, \Theta_{k-1})}P(\bold{\tilde{X}^{(k)}_{i}}=x_{i}\big | \bold{\tilde{X}^{(k)}_{j}}=x_{j}, \Theta_{k-1})\\
&=\sum_{x_{i}}{P(y^{k}_{ij} \big| \bold{\tilde{X}^{(k)}_{j}}=x_{j},\bold{\tilde{X}^{(k)}_{i}}=x_{i})P(\bold{\tilde{X}^{(k)}_{i}}=x_{i}\big | \Theta_{k-1})}. 
\end{split}
\end{align}
In (\ref{iteration5}), we use the assumption that $\bold{\tilde{X}^{(k)}_{i}} \independent (\bold{\tilde{X}^{(k)}_{j}}\big| \Theta_{k-1})$ and $y^{k}_{ij} \independent \big(\Theta_{k-1}\big| \tilde{X}^{(k)}_{i},  \tilde{X}^{(k)}_{j}\big)$. The first assumption results from the fact that given all the previous aggregated information in the network, update on location of $S_{i}$ at each time step is independent of that of $S_{j}$. Whereas the second assumption indicates that given the most recent updates on $S_{i}$ and $S_{j}$, the path loss between $S_{i}$ and $S_{j}$ is independent of the previously aggregated data in the network.

Combining (\ref{iteration4}) and (\ref{iteration5}) yields 
\begin{equation}
\label{iteration6}
P(Y^{(k)}_{j} \big | {\bold{\tilde{X}^{(k)}_{j}}}=x_{j}, \Theta_{k-1})=\prod_{i \in N_{j}}{\sum_{x_{i}}{\big [P(y^{k}_{ij} \big| \bold{\tilde{X}^{(k)}_{j}}=x_{j},\bold{\tilde{X}^{(k)}_{i}}=x_{i})P(\bold{\tilde{X}^{(k)}_{i}}=x_{i}\big | \Theta^{(k-1)}_{j} )\big ]}}. 
\end{equation}
Finally combining (\ref{iteration3}) and (\ref{iteration6}) completes the recursive update, 
\begin{align}
\begin{split}
\label{iteration7}
&P(\bold{\tilde{X}_{j}}=x_{j}\big| \Theta_{k})\propto P({\bold{\tilde{X}_{j}}}=x_{j} \big| \Theta_{k-1}) \times \\
&\prod_{i \in N_{j}}{\sum_{x_{i}}{\big [P(y^{k}_{ij} \big| \bold{\tilde{X}^{(k)}_{j}}=x_{j},\bold{\tilde{X}^{(k)}_{i}}=x_{i})P(\bold{\tilde{X}^{(k)}_{i}}=x_{i}\big | \Theta_{k-1})\big ]}}. 
\end{split}
\end{align}
This means in order to update posterior of $S_{j}$ after observation of new samples collected from $S_{i}$, we need to know priors of $S_{i}$ and $S_{j}$ in addition to channel information $P(y^{k}_{ij} \big| \bold{\tilde{X}^{(k)}_{j}}=x_{j},\bold{\tilde{X}^{(k)}_{i}}=x_{i})$. With respect to total number of nodes $n$, the algorithm has computational and communication complexity of $O(n)$ and $O(1)$ per node which renders the algorithm scalable. The computational complexity is the same as BP-based techniques whereas communication overhead which makes up most of power consumption in WSNs is significantly lower since each node only communicates with its single-hop neighbours.  

\subsection{Localization Algorithm Compatible with Wireless Sensor Networks}
\label{wsncompatible}
\noindent In this section, we proceed with a realization of the general case algorithm which is a more specific case of the proposed recursive solution in (\ref{iteration7}). Moreover we assume that at $k$-th time step, only $S_{k}$ does the multicasting and all connected nodes update their location posterior based on the observed path loss or mean of the path loss samples, i.e., $Y^{(k)}_{j}=y^{k}_{kj}$. This means each node is recipient of at most one sample at a single time step which guarantees compatibility with real world deployment of WSNs such as TDMA or carrier sense multiple access with collision avoidance (CSMA/CA)  where at each time slot, a node can listen to at most one neighbour node without interference. To be more specific, AODV which is the underlying routing protocol in ZigBee works based on flooding and multicasting route request (RREQ) packets and receiving routing reply (RREP) messages, hence our proposed localization algorithm can be integrated in a convenient and inexpensive manner. 

Off-the-shelf IEEE802.15.4 compliant modules such as TelosB, MICAz and Synapse modules give firmware engineers and designers the option to program them via Universal Serial Bus (USB), universal asynchronous receiver/transmitter (UART) ports or over-the-air (OTA). Even though MICAz and TelosB motes are widely used for academic and research purposes, Synapse modules which are equipped with light and fast network operating system, SNAP, and a more powerful microcontroller are more frequent for outdoor and industrial applications and better suited to more complex programming (with Python) and also mesh networking. In the next section, we use numerical examples to evaluate the performance of our algorithm based on radio characteristics of Synapse radio frequency (RF) modules. 
\begin{algorithm}[!h]
\begin{minipage}{16cm}
\textbf{Step 1}: Initialization (path loss model auto-tuning if required) \\
For $i =1,\ldots,n_{a}$  \, \, \, \, \, \, \, \textit{initializing landmarks locations}
\begin{itemize}
\item $P(\mathbf{\tilde{X}^{(0)}_{i}})  \sim \delta_{x_{i}}(x)$
\end{itemize}
For $i =n_{a}+1,\ldots,n_{t}$ \, \, \, \textit{initializing unknown nodes locations}
\begin{itemize}
\item $P(\mathbf{\tilde{X}^{(0)}}_{i})=  \frac{1}{m^{2}}$
\end{itemize}
\textbf{Step 2}: Landmarks advertise themselves to unknown nodes \\
For $i =1,\ldots,n_{a}$ \\
For $\forall j \in N_{i}$ \\
\begin{itemize}
\item $P(\mathbf{\tilde{X}^{(i)}_{j}}\big | \Theta_{i})= P(\mathbf{\tilde{X}^{(i-1)}_{j}}\big | \Theta_{i-1})P(y^{i}_{ij} \big|\mathbf{\tilde{X}^{(i-1)}_{j}}=x_{j},\mathbf{\bold{\tilde{X}^{(i-1)}_{i}}}=x_{i})$    
\item  Normalize $P(\mathbf{\tilde{X}^{(i)}_{j}}\big | \Theta_{i})$
\item Multicasting and updating with (\ref{iteration7}) continue till all unknown nodes are covered for each landmark advertisement. 
\end{itemize}
\textbf{Step 3}: A random node $S_{i}$ becomes source and multicasts RREQ packet   \\
\textbf{Step 4}: \\
For $j =n_{a}+1,\ldots,N$  \\
\begin{itemize}
\item  If $d_{ij}<d_{connectivity},$ \,  $j \neq i$ 
\begin{itemize}
\item Updating rule (\ref{iteration7})
\item  Normalization
\item $S_{j}$ forwards and multicasts the RREQ packet if hop count allows (AODV)
\end{itemize}
\item else
\begin{itemize}
\item $P(\bold{\tilde{X}^{(i)}_{j}}=x_{j}\big| \Theta_{i})=P(\bold{\tilde{X}^{(i-1)}_{j}}=x_{j}\big | \Theta_{i-1})$   \text{no change in location estimation}
\end{itemize}
\end{itemize}
While RREQ packet has not reached the landmark \\
$i \leftarrow \forall j \in N_{i}$ \\
Redo step 4 \\
\textbf{Step 5}: Landmarks return the RREP packet over the minimum hop route towards source   \\
For $\forall$ consecutive pairs of $(i,j)$ on landmark-source route \\
\begin{itemize}
\item $P(\mathbf{\tilde{X}^{(i)}_{j}}\big | \Theta_{i})= P(\mathbf{\tilde{X}^{(i-1)}_{j}}\big | \Theta_{i-1})P(y^{i}_{ij} \big|\mathbf{\tilde{X}^{(i-1)}_{j}}=x_{j},\mathbf{\bold{\tilde{X}^{(i-1)}_{i}}}=x_{i})$ \\
\item  Normalize $P(\mathbf{\tilde{X}^{(i)}_{j}}\big | \Theta_{i})$ \\
\end{itemize}
\begin{itemize}
\item else
\begin{itemize}
\item $P(\mathbf{\tilde{X}^{(i)}_{j}}\big | \Theta_{i})=P(\mathbf{\tilde{X}^{(i-1)}_{j}}\big |  \Theta_{i-1})$ \\
\end{itemize}
\end{itemize}
Go back to \textbf{Step 3} \\ 
\textbf{Step 5}: Decision making after M time steps  \\
For $j =n_{a}+1,\ldots,N$  \\
\begin{itemize}
\item $\tilde{x}_{j}=\underset{x_{j}}{\operatorname{argmax}}{[P(\mathbf{\tilde{X}^{(M)}_{j}}=x_{j}\big| \Theta_{N})]}.$
\end{itemize}
\end{minipage}
\caption{Localization Algorithm For Agricultural Environments}
\label{localizationalgorithm}
\end{algorithm}
\paragraph*{Quantization and Compression} There are limitations in terms of maximum payload size (102~bytes) which is imposed by underlying PHY and MAC layers. This limits us in terms of resolution of the exchanged pmf messages in the network and may prevent the localization algorithm from achieving the desired accuracy in large orchards. Therefore, there is a trade-off between localization accuracy and excessive power consumption in addition to delay which are caused by exchange of multiple packets between a pair of nodes for the sake of transferring the entire pmf message. Our simulations show that quantization and compression techniques are applicable so that pmf messages with more bins fit in a single packet. Discrete cosine transform (DCT), and 6-bit quantization help achieve compression ratio of up to 8/1 which translates to coverage of a 100~hectare (ha) orchard for high node density (7~nodes/ha) pest management (mating disruption) application. 

\begin{figure}[!htb]
\label{maoerrors}
\begin{subfigure}[b]{0.5\textwidth}
\includegraphics[width=\textwidth,height=0.8\textwidth]{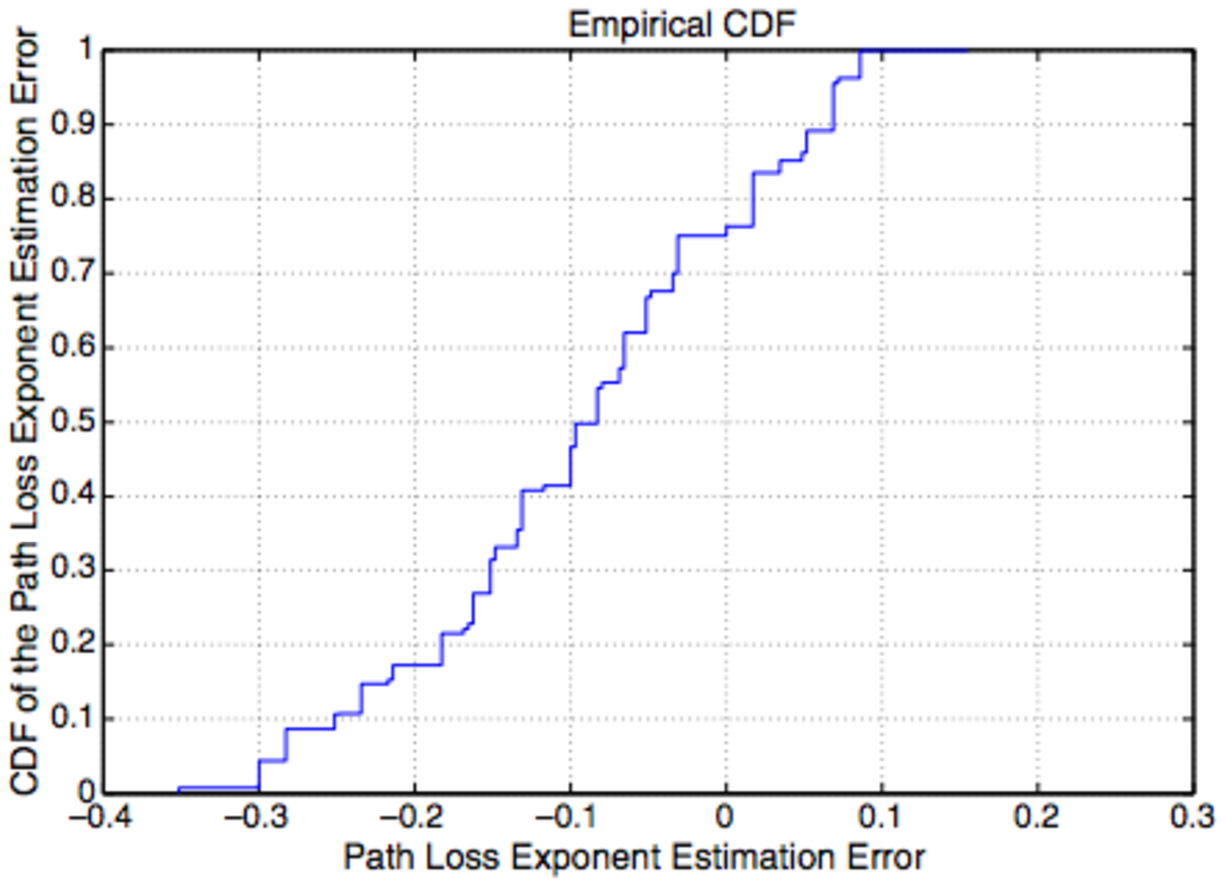}
\caption{path loss exponent estimation error}
 \label{exponentestimation} 
\end{subfigure}
\begin{subfigure}[b]{0.5\textwidth}
\includegraphics[width=\textwidth,height=0.78\textwidth]{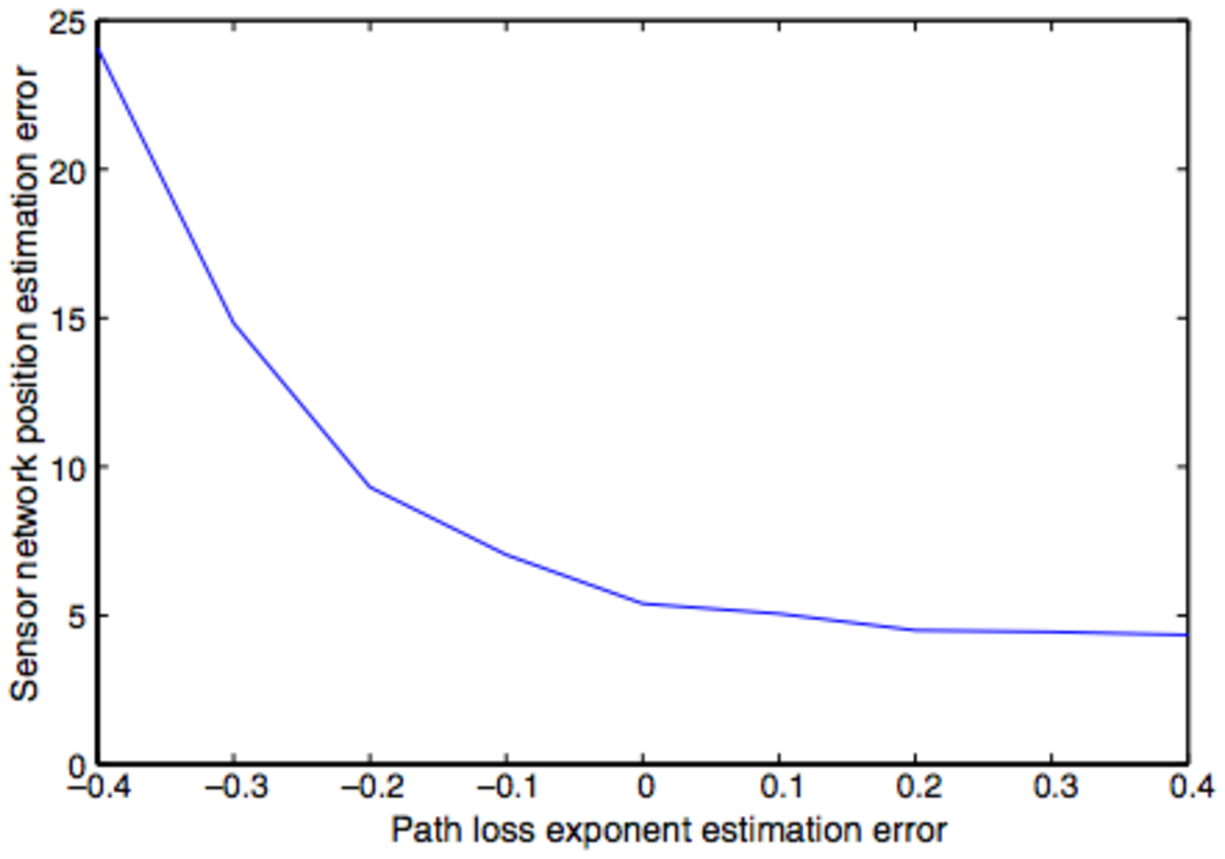}
\caption{Location estimation error}
 \label{distanceerror} 
\end{subfigure}   
\caption{Path loss exponent and localization estimation error claimed by Mao et al. \cite{PLEestimation} for randomly scattered nodes and landmarks in the corners.} 
\end{figure}

\paragraph*{Path Loss Model Auto-Tuning} So far we have assumed that there is a global awareness of path loss model among sensors, however this is not a realistic assumption due to remarkable changes during seasonal environmental variations. In \cite{PLEestimation}, Mao et al. proposed a path loss exponent estimation method based on Cayley--Menger determinant technique and pattern matching. The technique estimates path loss exponent with a high accuracy ($\approx \pm 0.2$) for the same landmark scenario that we have used in Section \ref{results}, i.e., landmarks deployed in the corners of the field, with estimation errors illustrated in Figures \ref{exponentestimation}, \ref{distanceerror}. Location estimation error could be tolerated for pest management applications for which the inter-node distance is 40~m-60~m.    

\paragraph*{Precision Agriculture Accuracy Requirements}
Coverage area of the sensors, spatial correlation of the measured features and required distance between actuators determine  inter-node distance for  deterministic grid WSN deployments. Further, inter-node distance could vary from 10~m for soil moisture \cite{Majone} and electrical conductivity \cite{farmingtools}, to coarser resolutions, 60~m for pH sensing \cite{mapquality} or mating disruption applications \cite{pheromone}. As will be seen in Section \ref{results}, our algorithm is mostly suited to pest management and mating disruption applications where tolerance for error which could result from the algorithm simplifying assumptions or mistuned path loss model.       
\section{Performance Evaluation of The Localization Algorithm}
\label{results}
\noindent In this section, we present the simulation results regarding performance of our localization scheme. We do the simulations for both random and deterministic (grid) deployment of WSN on a square field. We particularly use simulations to show that the average number of unknown nodes and landmarks each node connects to, affect the accuracy of the localization algorithm for a specific landmark arrangement. Hence, we define two parameters, so called \textit{average landmark degree} and \textit{average unknown node degree}. Let landmark and unknown node degree of an arbitrary node $S_{i}$ be the number of landmark and unknown nodes $S_{i}$ is connected to. Note that node degree in graph theory is strongly related to connectivity in the communications context. Further, average unknown node degree depends on deployment density and transmit power level of unknown nodes whereas transmit power of landmarks, location of the landmarks and number of them affect the landmark average degree. Different metrics have been used to evaluate performance of the localization algorithms \cite{QualityMetrics}. We use \rm{Twice the Distance Root Mean Square} (2DRMS) as the accuracy metric for our localization technique where \rm{2DRMS=r} means there is $95\%$ confidence that the location estimation would fall within a circle with radius $r$ around the actual node's location. Note that location estimation itself is a random variable due to random nature of path loss samples, and generating source node. This is due to event-driven data delivery model which is normally used for precision agriculture applications which means that a sensor transmits data only when a feature exceeds a predetermined threshold, hence message passing schedule is different after landmarks advertise themselves. The random nature of the problem makes 2DRMS a suitable accuracy metric.    

In this work, we do not concentrate on optimizing landmarks location however in the next section we explain the logic behind our adopted landmark arrangement. In the remainder of this section, first we explain the simulation setup and assumptions. We will then proceed with numerical examples to evaluate the performance of our algorithm. 

\begin{figure}[!htb]
\centering 
\includegraphics[width=4in]{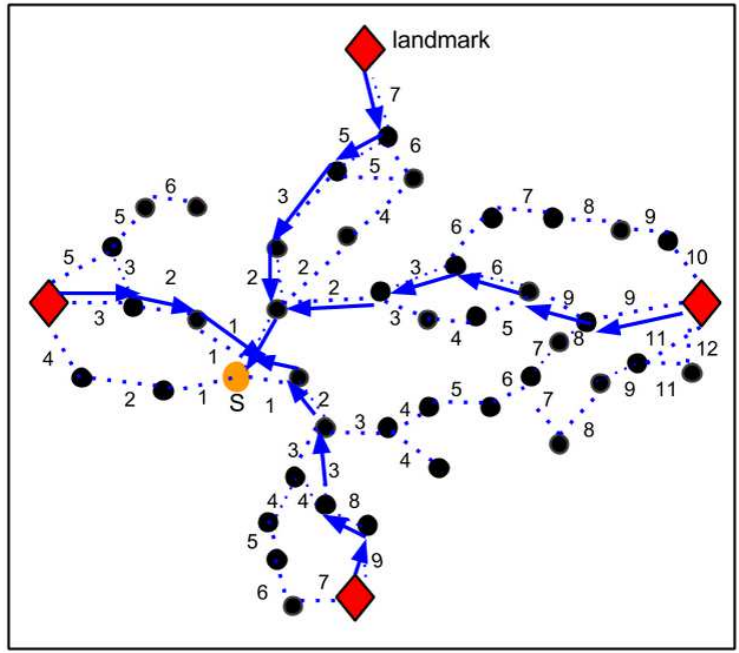} 
\caption{A schematic view of ZigBee route discovery phase with RREQ packet from source node (orange) being flooded in the network to reach landmarks (dotted paths) and RREP packets returning to the source node (solid arrows). Localization could be done both in conjunction with RREQ flooding or RREP return phase. Each number shows the number of times RREQ packet has been multicasted.}
\label{aodv}
\end{figure}

\subsection{Methodology}
\begin{figure}[!htb]
\begin{subfigure}[b]{0.5\textwidth}
\includegraphics[width=\textwidth,height=0.8\textwidth]{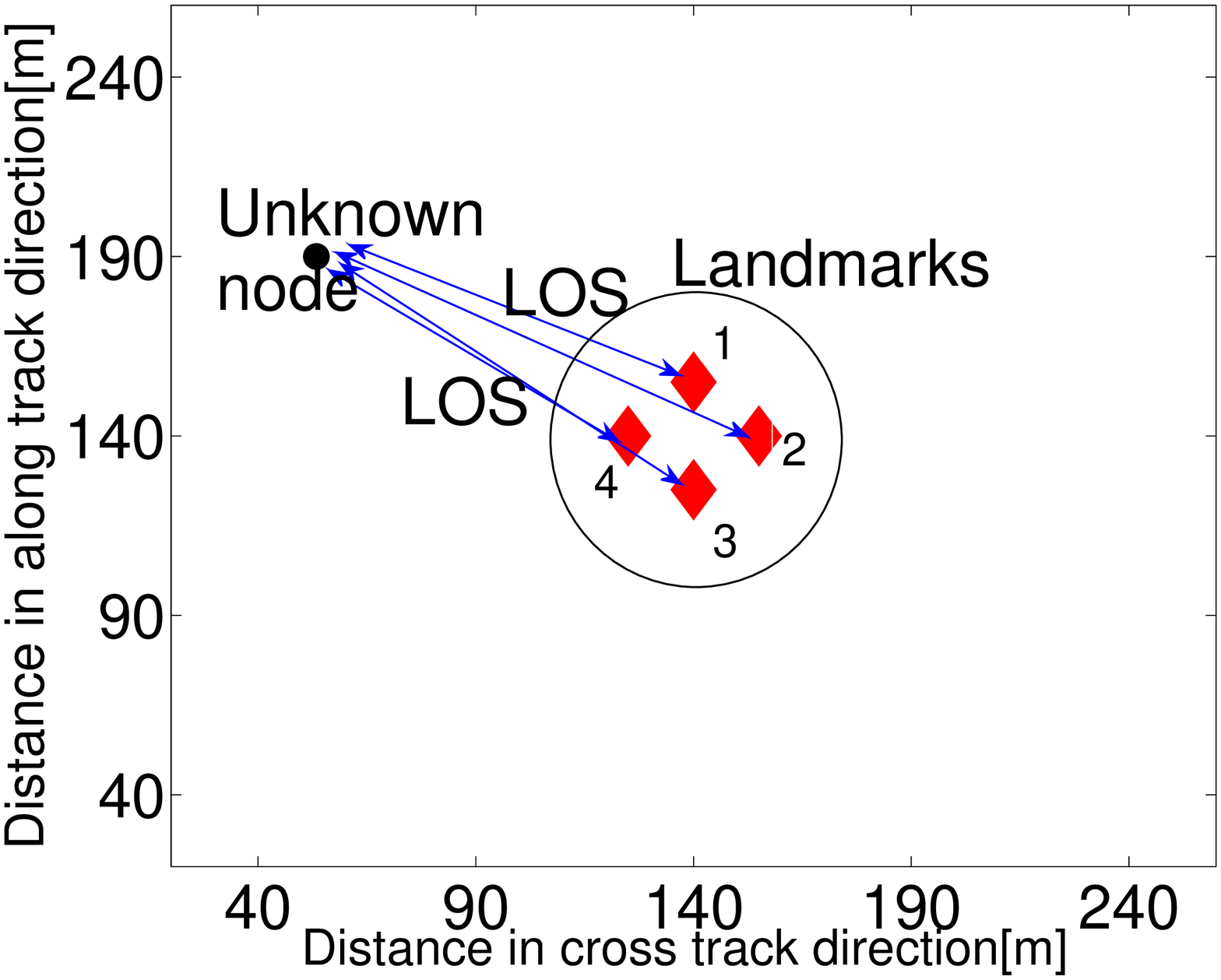}
\caption{Landmarks placed in the middle with every one of them having line of sight to the unknown node} \label{4los} 
\end{subfigure}
\begin{subfigure}[b]{0.5\textwidth}
\includegraphics[width=\textwidth,height=0.8\textwidth]{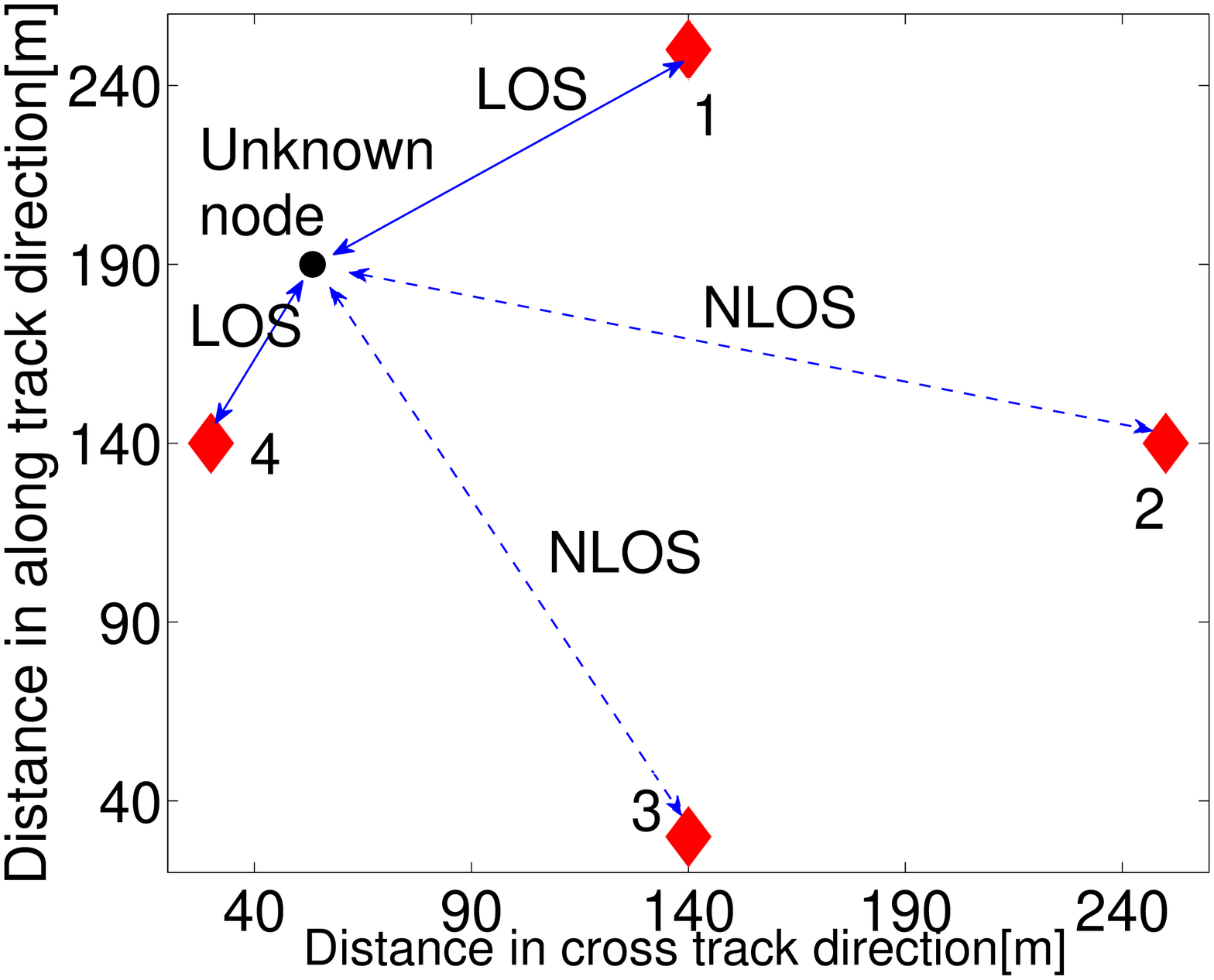}
\caption{Landmarks placed on the borders with only two of them having line of sight to the unknown node} \label{2los} 
\end{subfigure}
\caption{Two different landmark arrangements; The landmark arrangement in plot \ref{2los} provides more information about location of the unknown node despite having fewer nodes having line of sight to the unknown node}
\label{landmarkarrangement}
\end{figure}

\noindent In this section, it is first explained why we opt for placing landmarks in the corner or middle of border lines, and continue with justifying assumptions regarding adopted transmit power, orchard size and node density. For precision agriculture applications inside farms, gateways are placed on the corners and borders of the field, however in the following we provide some logic on why this helps towards the improvement of localization algorithm.  
\paragraph*{Landmark Arrangement}
\label{landmarklocationchap}
\noindent Even though placing landmarks close to each other and at the centre of the field yields a higher average landmark degree, the localization accuracy drops dramatically since their path loss behaviour has a very high correlation at a given direction and the path loss sample we collect from them is fairly close to each other at a specific point of reach. Moreover we place landmarks on the middle of borderlines or in the corners since the arrangement provides more information about unknown node's location. In Figure \ref{landmarkarrangement}, for a random unknown node location, it can be seen that having a more landmark degree does not necessarily result in a better location estimation. This is because distances in Figure \ref{4los} are fairly close to each other and given that a noisy estimation of them are made based on path loss samples, the location estimation will be far less accurate compared to the arrangement in Figure \ref{2los}. It can be easily shown that this scenario holds for most points on the field. Studying other landmark arrangements could be done accordingly, however we avoid to elaborate on it for the sake of space considerations and since it does not add to evaluation of the algorithm and is therefore beyond the scope of this work.  

\begin{figure}[!htb]
\begin{subfigure}[b]{0.5\textwidth}
\includegraphics[width=\textwidth,height=0.8\textwidth]{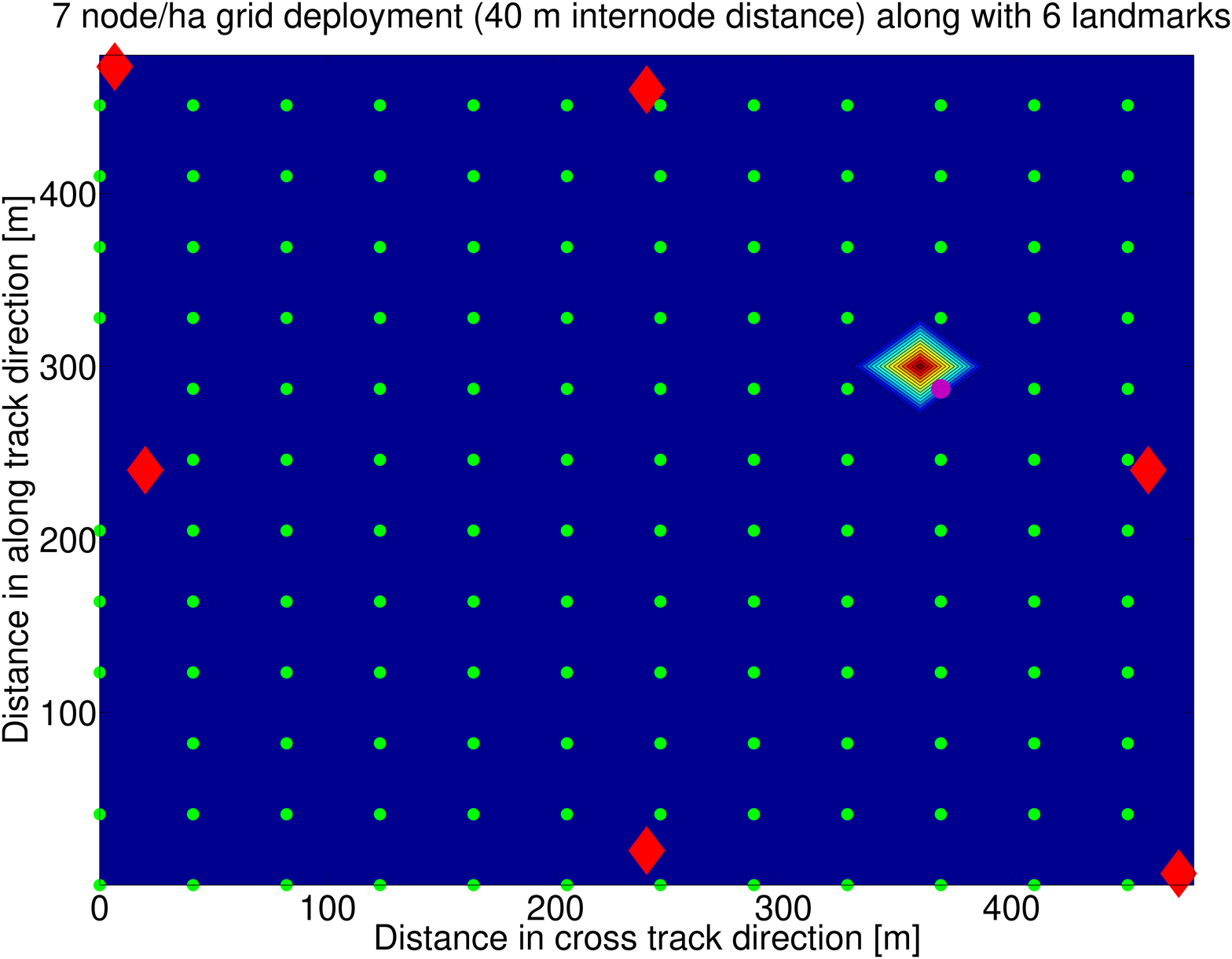}
\caption{6 landmarks} 
\label{heatmap1} 
\end{subfigure}
\begin{subfigure}[b]{0.5\textwidth}
\includegraphics[width=\textwidth,height=0.8\textwidth]{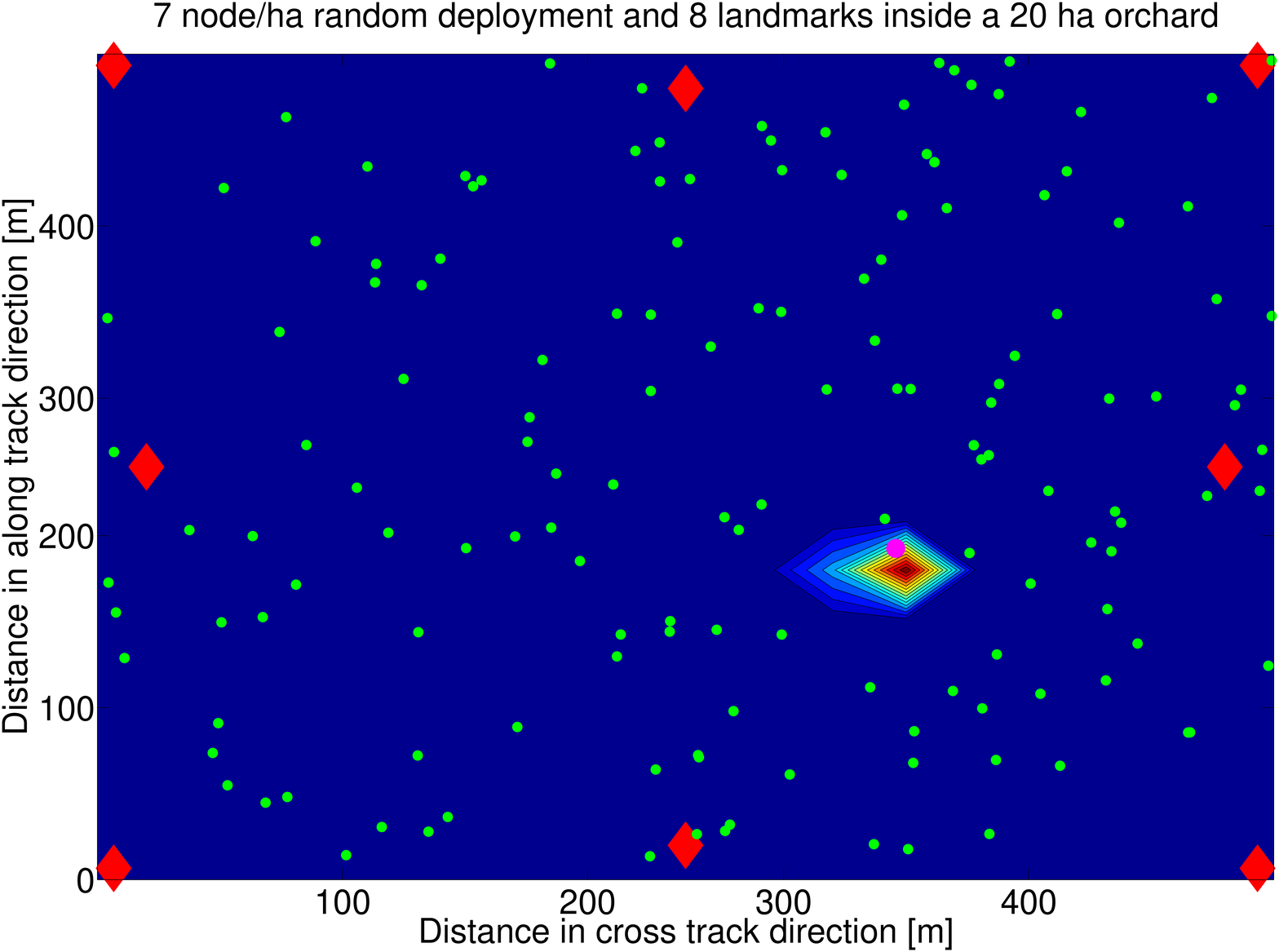}
\caption{8 landmarks} 
\label{heatmap2} 
\end{subfigure}
\caption{Two different landmark arrangements; unknown nodes and landmarks are demonstrated with green small dots and red large diamonds respectively. Location pmf for the designated unknown node (purple) is illustrated by heat map.}
\label{LandmarkSetup}
\end{figure}

\begin{figure}[!htb]
\centering 
\includegraphics[width=4in]{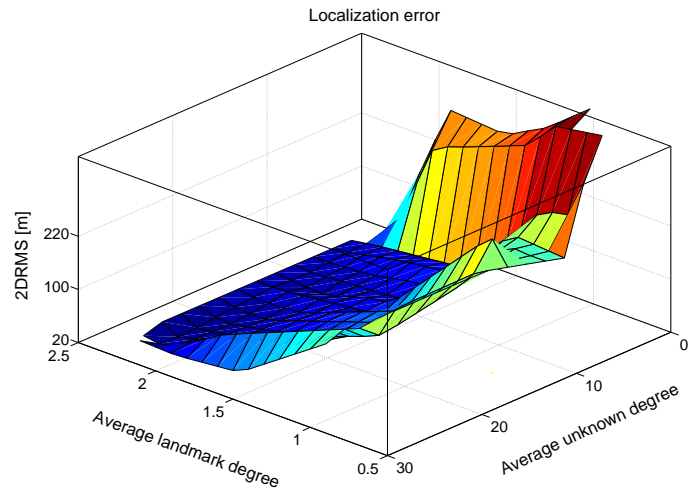} 
\caption{{\rm 2DRMS} with respect to average landmark and unknown node degree is depicted. Surface points are collected from all deployment scenarios}
\label{averagedegree1}
\end{figure}

\begin{figure}[!htb]
\centering 
\includegraphics[width=4in]{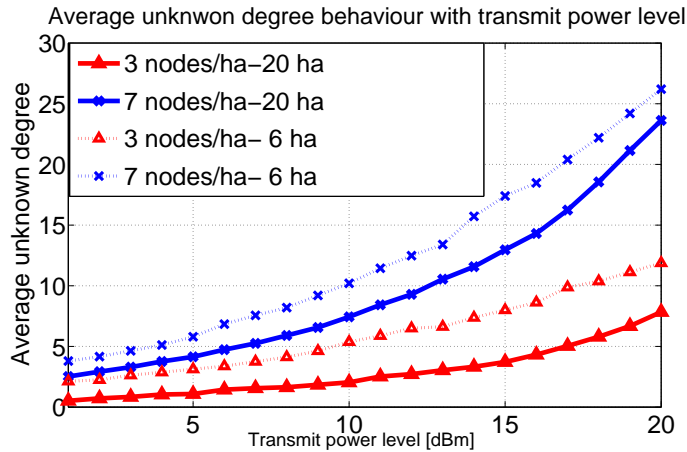} 
\caption{average unknown node degree of nodes with respect to transmit power level of the unknown nodes. Dotted and solid graphs represent deployment scenarios for 6~ha and 20~ha orchards respectively.}
\label{averagedegree2}
\end{figure}

\begin{figure}[!htp]
\begin{subfigure}[b]{0.5\textwidth}
\includegraphics[width=\textwidth,height=0.8\textwidth]{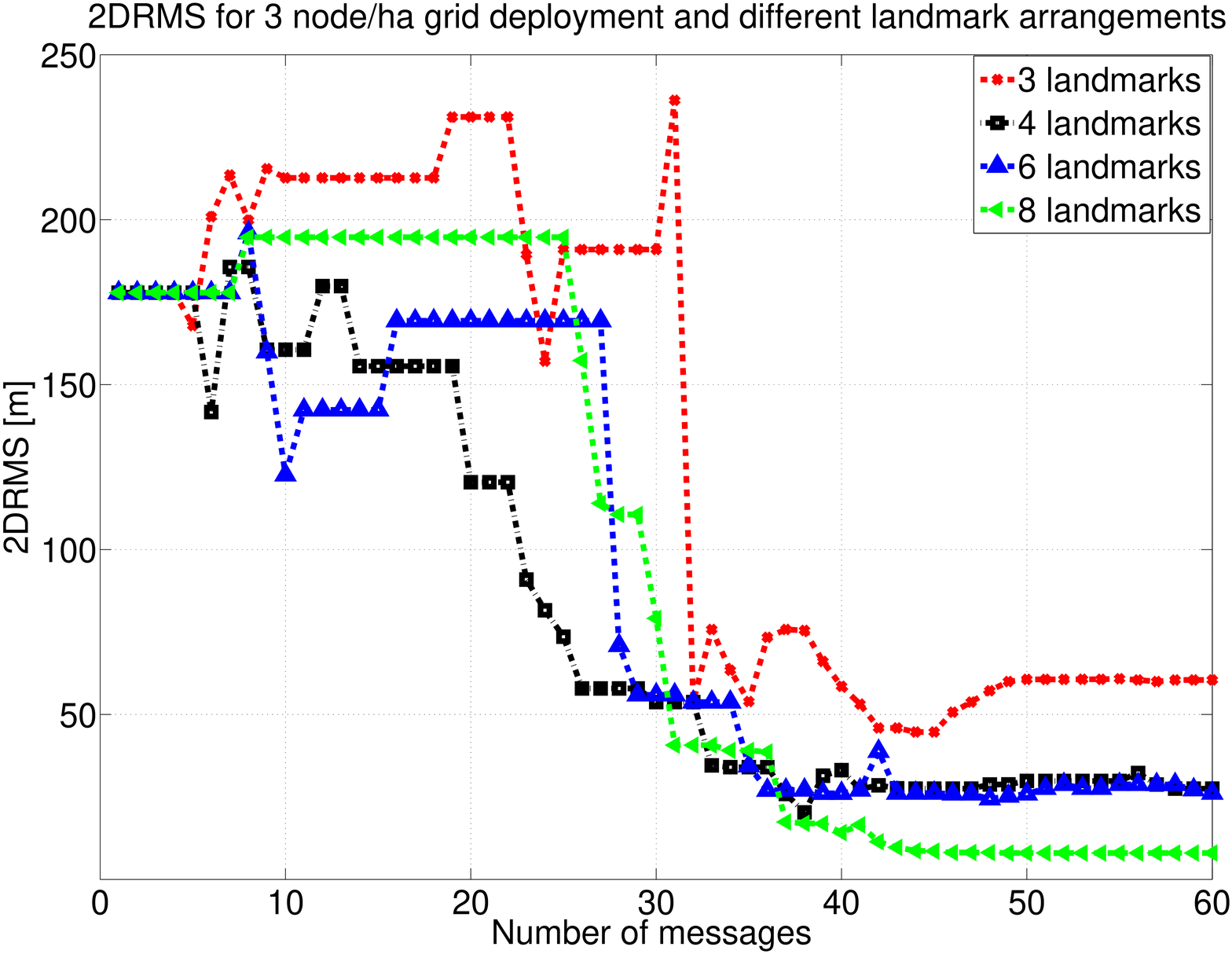}
\caption{Low density grid deployment} 
\label{error50grid} 
\end{subfigure}
\begin{subfigure}[b]{0.5\textwidth}
\includegraphics[width=\textwidth,height=0.8\textwidth]{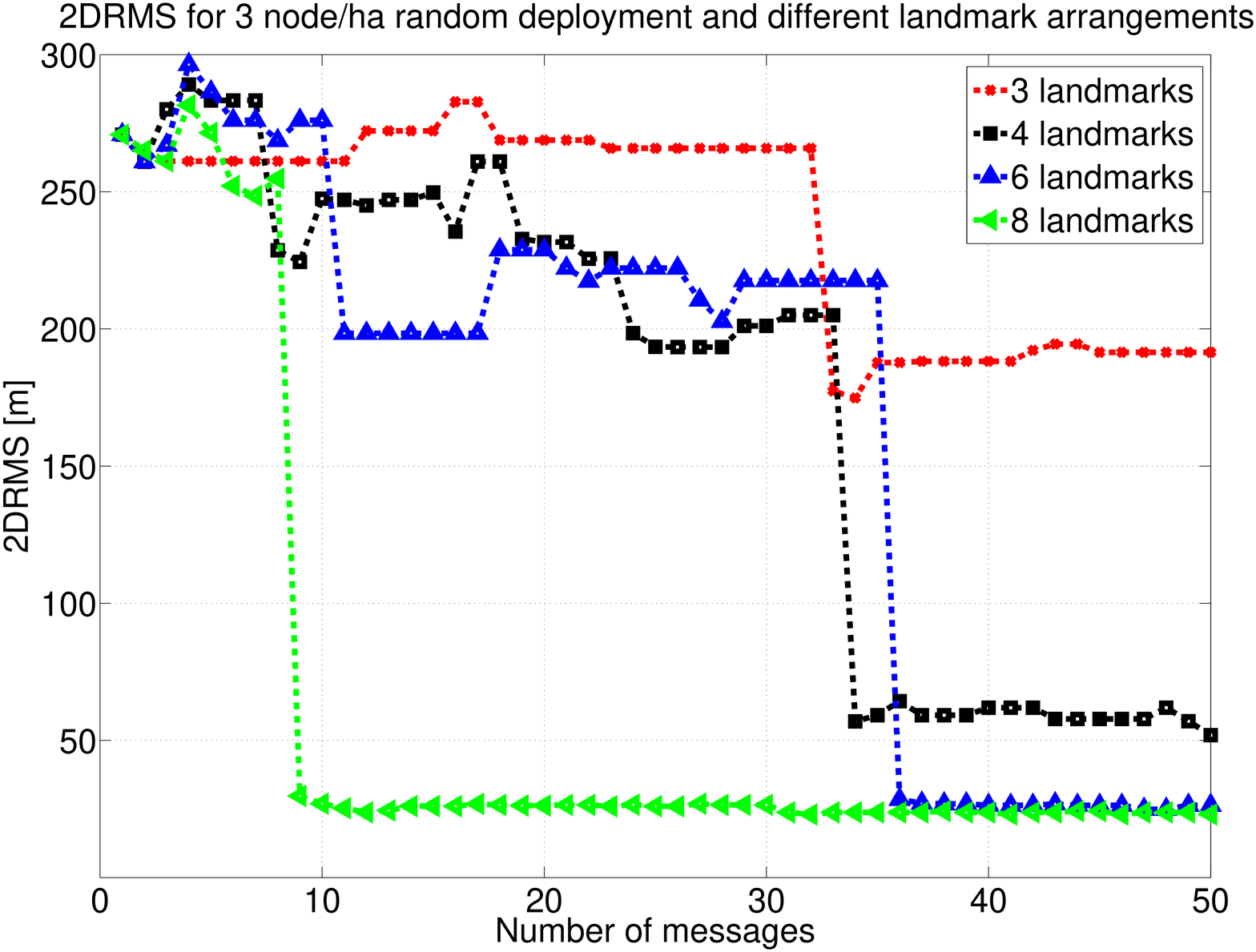}
\caption{Low density random deployment}
\label{error50rnd} 
\end{subfigure}
\begin{subfigure}[b]{0.5\textwidth}
\includegraphics[width=\textwidth,height=0.8\textwidth]{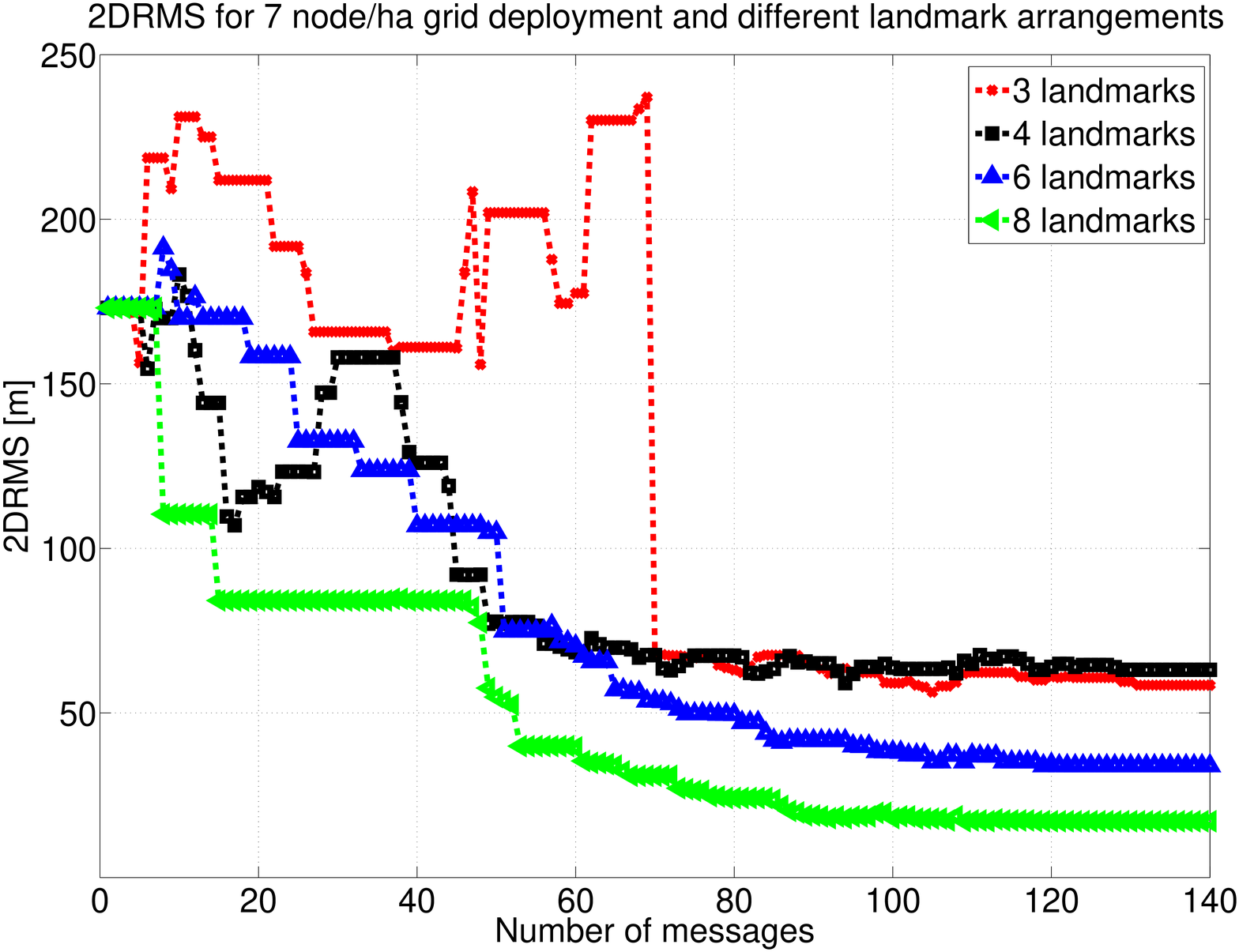}
\caption{High density grid deployment}
 \label{error150grid} 
\end{subfigure}
\begin{subfigure}[b]{0.5\textwidth}
\includegraphics[width=\textwidth,height=0.8\textwidth]{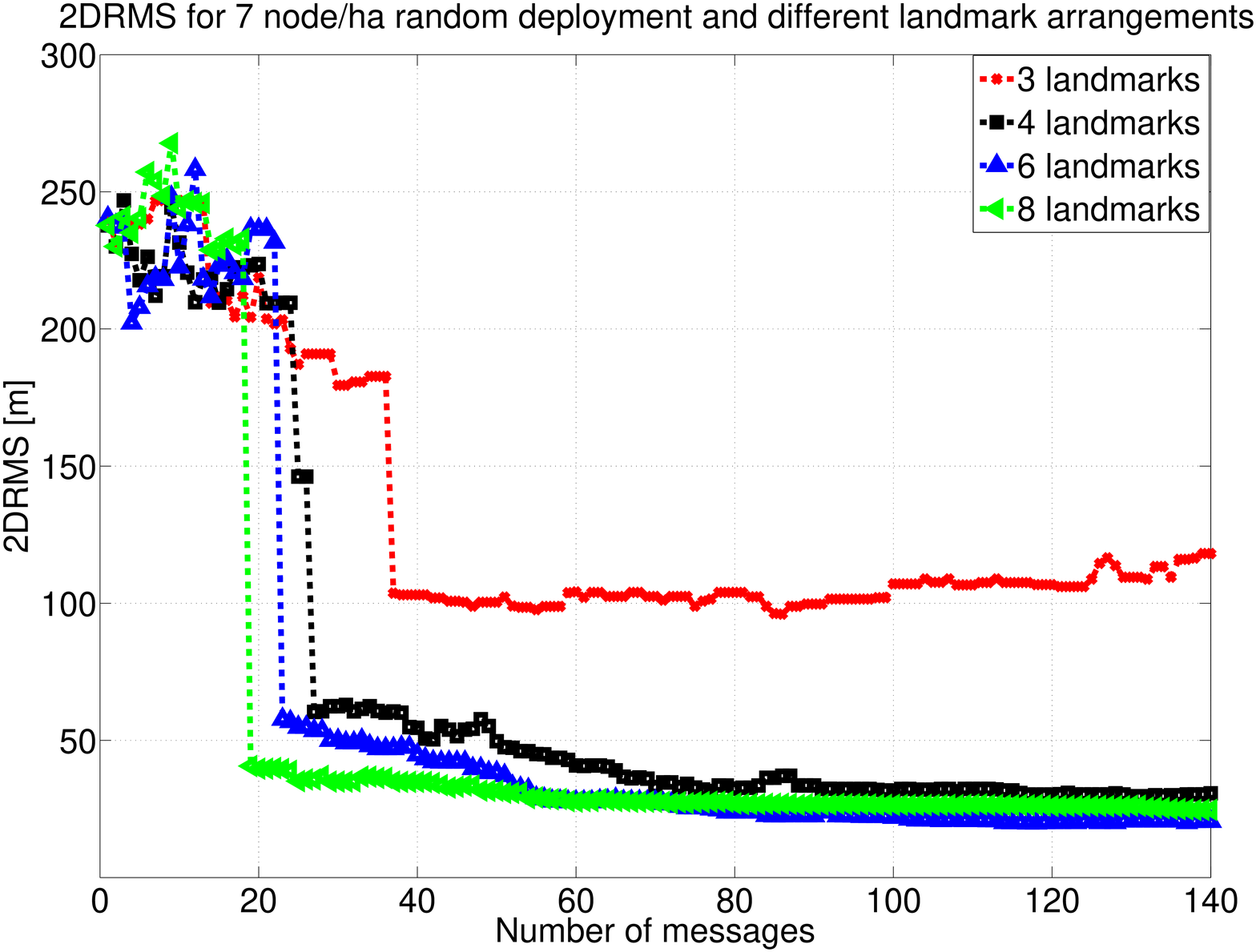}
\caption{High density random deployment}
 \label{error150rnd} 
\end{subfigure}
\caption{{\rm 2DRMS} for different node densities and landmark arrangements inside a 20~ha apple orchard; Low and high density grid deployments translate to 60~m and 40~m internode distance respectively and is well-suited to mating disruption.}
\label{2drmsbehaviour}
\end{figure}
\begin{figure}[!htp]
\begin{subfigure}[b]{0.5\textwidth}
\includegraphics[width=\textwidth,height=0.8\textwidth]{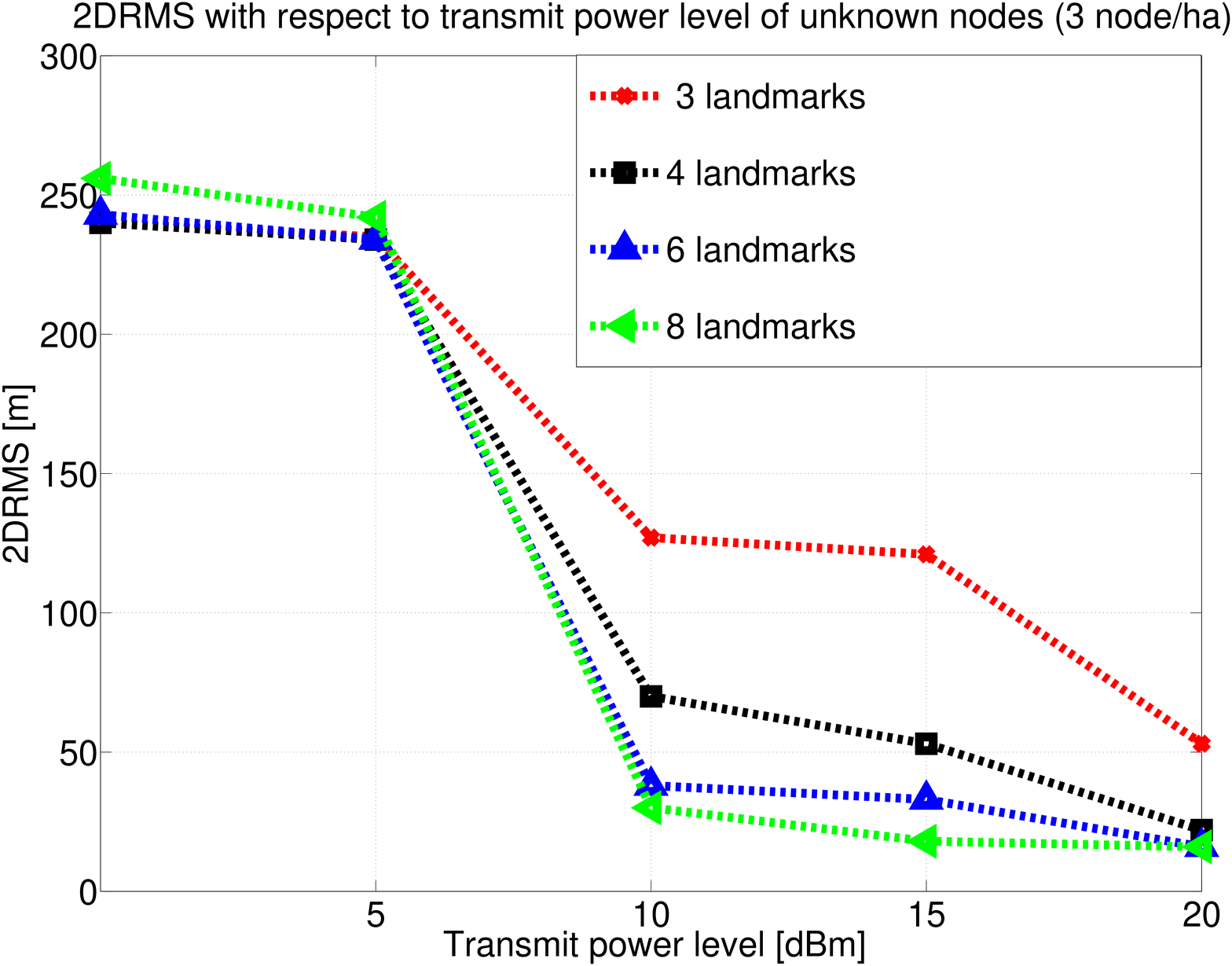}
\caption{} 
\label{loc2drmspower3n} 
\end{subfigure}
\begin{subfigure}[b]{0.5\textwidth}
\includegraphics[width=\textwidth,height=0.8\textwidth]{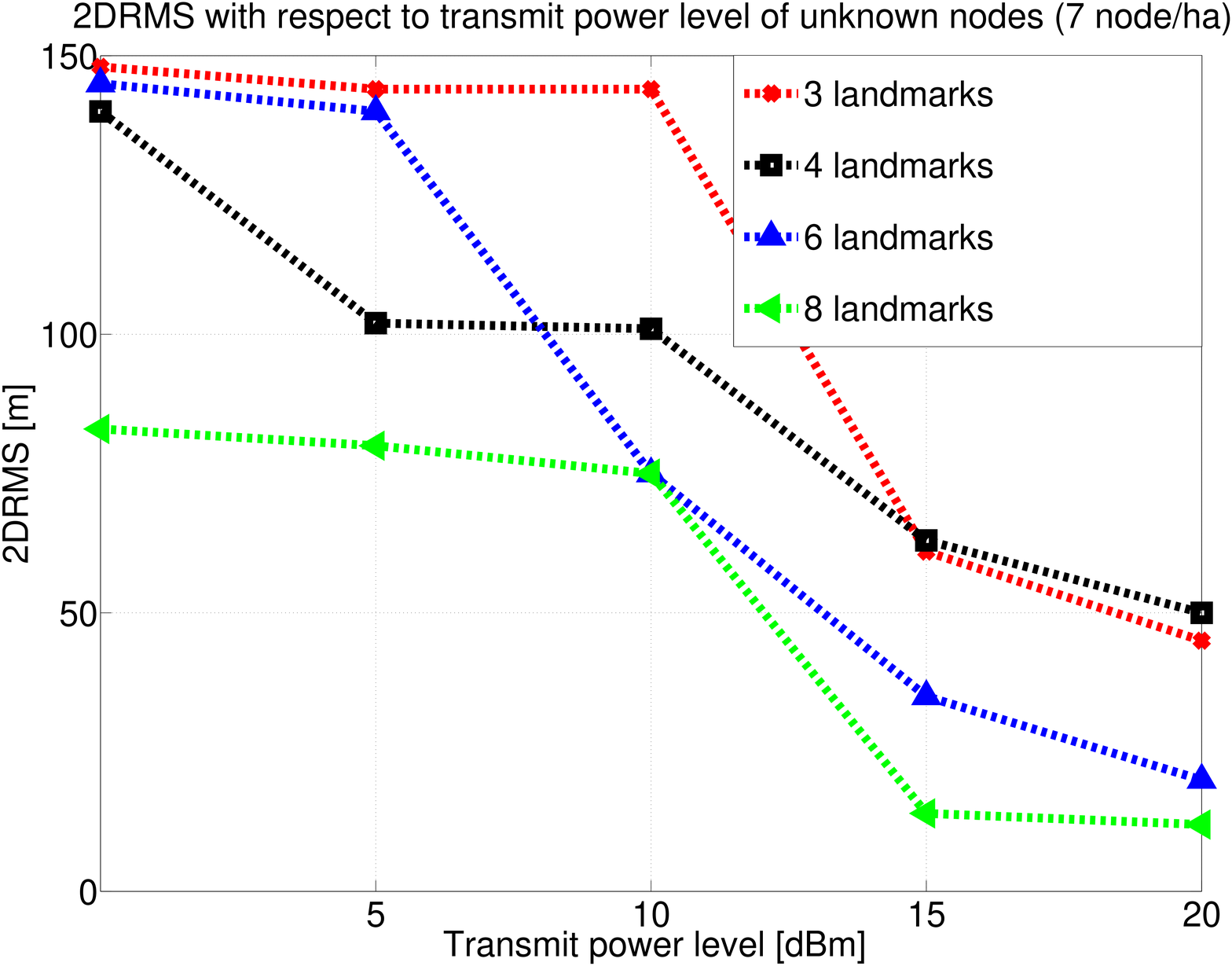}
\caption{}
\label{loc2drmspower7n} 
\end{subfigure}
\caption{{\rm 2DRMS} variations with transmit power level $P_{tx}$; different scenarios in terms of node density and number of landmarks inside a 20~ha apple orchard are illustrated. The 2 landmark scenario is excluded for the sake of clarity and lack of space since it achieves a fairly low accuracy. Increasing node density helps towards achieving low \rm{2DRMS} with lower number of landmarks and transmit power}
\label{2drmsVstxpower}
\end{figure}

\begin{table}[!htp]%
\caption{Deployment Scenarios}
\label{simulationsetup}\centering %
\rowcolors{1}{}{lightgray}
\begin {tabular}{cc}
\hline
Orchard size & 6~ha, 20~ha \\
Node density (nodes per hectare)& 3, 7   \\ 
Node arrangement &  Grid  \\
Transmit power level of unknown nodes& 0-15~dBm  \\ 
Transmit power level of landmarks& 15~dBm  \\   
Transmit power increment step & 1~dB \\
Receiver sensitivity for PER=1\% & -103~dBm \\
Grid cell dimension & 30~m \\
 Number of landmarks& 2,3,4,6 and 8    \\  
Location of landmarks & borders and corners  \\
Landmark degree (6~ha orchard) & varying from 1.78 to 6.3    \\
Landmark degree (20~ha orchard) & varying from 0.8  to 2.18   \\
Maximum transmission distance for below canopy mode & 120~m  \\
Maximum transmission distance for above canopy mode & 220~m
\end{tabular} 
\end{table}
\paragraph*{Deployment Scenarios and Assumptions}
\label{simulationsetupchap}
\noindent In our simulation setup which is summarized in Table \ref{simulationsetup}, we adopt two different orchard sizes of $6$ and 20 hectares (ha) with nodes randomly scattered inside the field at two different densities, 3~nodes/ha, and 7~nodes/ha. As discussed in Section \ref{IterativeAlgorithm}, these are the densities used for pest management applications and translate to 60~m and 40~m inter-node distance for grid deployment respectively. Grid cell dimension is chosen to be 30~m so that both these densities could be covered.  The average size of an apple orchard varies from 1 to 20~ha in different regions,  whereas the average size in Canada and the United States is approximately 6~ha and 20~ha respectively according to the United States Department of Agriculture \cite{applefacts}. Node density and type of deployed RF modules may vary based on the precision agriculture application and required sampling range \cite{Deployment1}. We also adopt four landmark arrangements with transmit power level of unknown nodes varying from $0$ to $+15$~dBm, receiver sensitivity for packet error rate (PER) to be $-103$~dBm, whereas the communication between landmark and nodes occurs at maximum transmit power ($+15$~dBm). Variation of landmark degree for different number of landmarks and orchard sizes is also expressed in Table \ref{simulationsetup} which are based on the assumption that Synapse RF200 modules are used \cite{synapse}.
                         
We also assume that landmarks (gateways) and unknown nodes (sensors) are mounted above and below canopy level respectively. We call $S_{i}$ and $S_{j}$ connected, $d_{ij}<d_{connectivity}$, in case the probability of RSSI falling below receiver sensitivity is below 1\% or connectivity probability is above 99\%. This maximum transmission distance is calculated based on our measurement-based path loss model summarized in Table \ref{pathlosstable}. In Table \ref{simulationsetup}, we have tabulated the transmission distance of Synapse RF200 module at its maximum transmit power so that connectivity requirement is met \cite{synapse}. In the next section we evaluate the performance of our algorithm. 
\subsection{Results}
\label{numericalexamples}
 \noindent In this section, we study the localization error of our algorithm for different simulation scenarios. In Figure \ref{LandmarkSetup}, two landmark arrangements, 6 and 8, along with 150 deterministically and randomly scattered sensors and maximum transmit power are illustrated. Location distribution of one designated node (purple node)  after the algorithm converges is illustrated. 
  
In Figure \ref{averagedegree1}, we illustrate the behaviour of \rm {2DRMS} with respect to average landmark and unknown node degree. As can be seen in the surface plot, error drops dramatically with average unknown node degree increasing. Further, even for a low average landmark degrees, $\approx 1.5$, an approximate average unknown node degree of  $8$ yields the desired {\rm 2DRMS}  ($\approx 20~m$). In Figure \ref{averagedegree2}, we demonstrate how average unknown node degree increases with transmit power level of unknown nodes in different simulation setups. These two figures provide an insight on how algorithm works with different transmit power levels. 

In Figure \ref{2drmsbehaviour}, \rm{2DRMS} behaviour for different simulation setups during course of the algorithm is demonstrated which shows that the algorithm converges after a few messages are multicasted in the network. As explained in Algorithm \ref{localizationalgorithm}, the procedure starts with landmarks advertising themselves to the entire network. This significantly helps towards faster convergence of the algorithm since one-hop neighbours of landmarks achieve a narrower pmf estimation at the first round. As could be seen in the Figure, generally 6 and 8 landmark/gateway scenarios meet the accuracy requirement for pest management, however in order to make the algorithm work for soil moisture sensing, number of landmarks or their maximum transmit power needs to increase. In other words, our simulations showed that a finer pmf resolution does not affect the accuracy in case cell dimension already supports the application in terms of inter-node distance. We also observed that the total number of messages needed for algorithm to converge grows slower than O(n) which is a promising aspect from the scalability stand of view. Moreover, in spanning tree variants of BP-based techniques, at least O(n) messages are required to make the spanning tree and after that every sensor needs to do a multicast at each iteration with algorithm taking anywhere between 1 to 3 iterations to converge. This means our algorithm is faster and consumes less communication energy to converge at the expense of accuracy.

 In Figure \ref{2drmsVstxpower}, localization error for a 20~ha orchard, 40~m and 60~m inter-node distances, with respect to transmit power level is depicted. Node density has higher influence at low transmit power levels which is compatible with our observations from Figure \ref{averagedegree1}. Once transmit power increases, at a fixed landmark degree, average unknown node degree exceeds the required threshold and error drops to minimum. Based on the work in \cite{PLEestimation} and our simulations, the algorithm meets pest management (mating disruption) requirements with acceptable probability (above 90\%) inside a 20~ha orchard with 8 landmarks and all unknown nodes running on Synapse RF200 modules, however a different transceiver module may demand for different landmark setups since the maximum transmit power level would be different. More landmarks are needed in larger orchards in order to meet the average landmark degree.   
 
\section{Conclusion}
 \label{conclusion}
\noindent Connectivity to landmarks in static WSNs deployed in large agricultural environments such as farms and orchards is limited due to excessive path loss and large size of the field. Besides, large number of nodes in the field and nature of higher layer communication algorithms in terms of transmit power and multicasting make connectivity graph for these WSNs very loopy. Most existing localization algorithms are ill-suited for use in such environments because they are overly complex, susceptible to loopy connectivity graphs, and incapable of real time updates, i.e., all the inter-node distance estimations must be completed before the algorithm runs.

\noindent Our scalable RSSI-based localization algorithm overcomes these limitations by: 
\begin{enumerate}
\item using only local distance estimates with respect to neighbouring nodes, 
\item a small number of landmarks compared to total number of nodes,   
\item adopting coarser or finer grid of the field based on the application and available processing power at microcontroller of the transceiver modules and desired localization accuracy for a specific precision agriculture application.
\end{enumerate}
The algorithm uses a Bayesian model for information aggregation to achieve scalable communication and computational complexity with respect to the number of nodes. The computational burden of the algorithm is divided between nodes and time steps. Besides, the algorithm could be stopped at any time step to carry out the decision making on the location of nodes.

\noindent The main strength of our localization algorithm is its compatibility with realistic deployment scenarios of
WSNs and the low communication overhead it adds to the already deployed routing protocols. Further,
the route discovery phase of ad hoc on-demand distance vector (AODV) routing protocols, e.g., ZigBee and
similar schemes, work based on flooding and multicasting route request (RREQ) packets; hence our
proposed localization algorithm can be integrated in a convenient and inexpensive manner.

\section{Acknowledgement}
We would like to greatly thank SemiosBio, Vancouver-based startup company specializing in precision agriculture, for generously supporting and funding our measurement campaigns in addition to granting us access to Dawson orchard at Keremeos, BC. SemiosBio's need for a localization algorithm which could run on Synapse transceiver modules to address mating disruption application was a significant inspiration for this work. We would also like to thank UBC radio science lab (RSL) 2012 and 2013 summer students for their hard work in terms of preparation of the measurement campaign setup and helping us conduct the measurements throughout warm summer seasons in Okanagan, British Columbia.        
\bibliographystyle{IEEEtran}
\bibliography{references}
\end{document}